\documentclass[aps,pre,twocolumn,superscriptaddress,10pt]{revtex4}
\usepackage[dvips]{graphics}
\usepackage{graphicx}
\usepackage{amsfonts}
\usepackage{amssymb}
\usepackage{amsmath}
\usepackage{subfigure}

\begin{document}

\title{\textbf{Polarized States and Domain Walls in Spinor Bose-Einstein
Condensates}}
\author{H. E. Nistazakis}
\affiliation{Department of Physics, University of Athens, Panepistimiopolis, Zografos,
Athens 15784, Greece }
\author{D. J.\ Frantzeskakis}
\affiliation{Department of Physics, University of Athens, Panepistimiopolis, Zografos,
Athens 15784, Greece }
\author{P. G.\ Kevrekidis}
\affiliation{Department of Mathematics and Statistics, University of Massachusetts,
Amherst MA 01003-4515, USA}
\author{B. A.\ Malomed}
\affiliation{Department of Interdisciplinary Studies, Faculty of Engineering, Tel Aviv
University, Tel Aviv 69978, Israel}
\author{R.\ Carretero-Gonz\'alez}
\affiliation{Nonlinear Dynamical Systems Group, Department of Mathematics and Statistics,
and Computational Science Research Center, San Diego State University, San
Diego CA, 92182-7720, USA}
\author{A. R. Bishop}
\affiliation{Theoretical Division and Center for 
Nonlinear Studies, Los 
Alamos National Laboratory, Los Alamos, New Mexico 87545, USA}

\begin{abstract}
We study spin-polarized states and their stability in anti-ferromagnetic
states of spinor ($F=1$) quasi-one-dimensional Bose-Einstein condensates.
Using analytical approximations and numerical methods, we find various types
of polarized states, including: patterns of the Thomas-Fermi type; 
structures with a pulse-shape in one component inducing a hole in
the other components; states with holes in all three components; and domain
walls. A Bogoliubov-de Gennes analysis reveals that families of these
states contain intervals of a weak oscillatory instability, except for the domain
walls, which are always stable. The development of the instabilities is
examined by means of direct numerical simulations.
\end{abstract}

\maketitle

\section{Introduction}

The development of the far-off-resonant optical techniques for trapping
ultracold atomic gases has opened new directions in the studies of
Bose-Einstein condensates (BECs). Atoms can be confined regardless
of their hyperfine (spin) state, thus avoiding freezing the atom's spin
degree of freedom, making it available for the study of interesting spin dynamics \cite{ket0}. One of the 
major achievements in this direction was the experimental creation of 
\textit{spinor} BECs \cite{ket1,cahn}. The spinor condensate formed by atoms
with spin $F$ is described by a $(2F+1)$-component macroscopic (mean-field)
wave function, which gives rise to various phenomena that are not present in
single-component BECs, including formation of spin domains \cite{spindomain}%
, spin textures \cite{spintext}, and multi-component (vectorial) solitons of
bright \cite{wad1,boris,zh}, dark \cite{wad2}, gap \cite{ofyspin} and
bright-dark \cite{ourspin} types.

Generally, the dynamics of the spinor $F=1$ BEC is spin-mixing \cite{sm}.
However, there also exist non-mixing, or \textit{spin-polarized }states of
the system, which are stationary solutions of the corresponding system of
Gross-Pitaevskii (GP) equations \cite{ho,yi}. The stability of such
polarized states is different for the two distinct types of the $F=1$
condensates, namely the \textit{ferromagnetic }(FM), such as in $^{87}$Rb,
and \textit{polar }(alias\textit{\ anti-ferromagnetic}, AFM), such as $^{23}$%
Na, ones, where the spin-dependent interactions are, respectively,
attractive and repulsive. Accordingly, as demonstrated in Ref. \cite{ofy2}
(see also Ref. \cite{boris}), spin-polarized states are modulationally
stable/unstable in the AFM/FM condensates.

In this work, we focus on AFM spinor condensates, and study, in
particular, spin-polarized states of the spinor BEC of $^{23}$Na
atoms. Assuming that this spinor system is confined in a strongly
anisotropic trap, we first present the respective system of three
coupled quasi one-dimensional (1D) GP equations. Then, employing
the so-called single-mode approximation \cite{sm,ofy2}, we use
analytical and numerical methods to find the spin-polarized states
of the system and study their stability via the
Bogoliubov-de Gennes (BdG) equations (i.e., the linearization of
the GP equations). The simplest possible form of these states are the stable 
Thomas-Fermi (TF)\ configurations (for each hyperfine component).
We also present other spin-polarized 
states, including ones in which one component maintains a
pulse-like shape, inducing a hole in the other components, and
structures with holes in all the three components. All these
states feature instability regions too, but with values of the
normalized instability growth rate up to $\sim 10^{-3}$.
Development of the oscillatory instabilities is examined by means
of direct simulations. It is found that the structures with a hole
in one component get weakly deformed, while the states with
holes in all the three components suffer stronger deformations. If the
three components are initially spatially separated, being confined
in different harmonic traps, but are put in a single trap
afterwards, spin domain-wall (DW) patterns are formed. A family of
the DW solutions exists and is fully stable if, for a fixed value
of the trap's strength, the chemical potential (or the number of
atoms) exceeds a certain critical value.

The paper is organized as follows: Section II presents the model. Sections
III is dealing with TF states. In Sections IV and V we examine structures
with one and multiple holes, respectively (including their stability).
Finally, conclusions are presented in Section VI.

\section{The model and setup}

At sufficiently low temperatures, and in the framework of the mean-field
approach, a spinor BEC with $F=1$ is described by a vectorial order
parameter, $\mathbf{\Psi }(\mathbf{r},t)=[\Psi _{-1}(\mathbf{r},t),\Psi _{0}(%
\mathbf{r},t),\Psi _{+1}(\mathbf{r},t)]^{T}$, with the different fields 
corresponding to three values of the vertical component of the spin, $%
m_{F}=-1,0,+1$. Assuming that this condensate is loaded into a strongly
anisotropic trap, with holding frequencies $\omega _{x}\ll \omega _{\perp }$%
, we assume, as usual, that the wave functions are separable, $\Psi _{0,\pm
1}=\psi _{0,\pm 1}(x)\psi _{_{\perp }}(y,z)$, where the transverse
components $\psi _{\perp }(y,z)$ represent the ground state of the
respective harmonic oscillator. Then, following the standard approach
\cite{gpe1d} of averaging the coupled 3D GP equations for the three components
in the transverse plane $\left( y,z\right) $, leads to the system of
coupled 1D equations for the longitudinal components of the wave functions
(see Refs. \cite{wad1,wad2,boris,ofyspin,ourspin}):
\begin{eqnarray}
i\hbar \partial _{t}\psi _{\pm 1} &=&\hat{H}_{\mathrm{si}}\psi _{\pm
1}+c_{2}^{(\mathrm{1D})}(|\psi _{\pm 1}|^{2}+|\psi _{0}|^{2}-|\psi _{\mp
1}|^{2})\psi _{\pm 1}  \notag \\
&+&c_{2}^{(\mathrm{1D})}\psi _{0}^{2}\psi _{\mp 1}^{\star },  \label{mgp1} \\
i\hbar \partial _{t}\psi _{0} &=&\hat{H}_{\mathrm{si}}\psi_{0}+c_{2}^{(%
\mathrm{1D})}(|\psi _{-1}|^{2}+|\psi _{+1}|^{2})\psi _{0}  \notag \\
&+&2c_{2}^{(\mathrm{1D})}\psi _{-1}\psi _{0}^{\star }\psi _{+1},
\label{mgp2}
\end{eqnarray}%
where star denotes complex conjugate, and $\hat{H}_{\mathrm{si}}\equiv
-(\hbar ^{2}/2m)\partial _{x}^{2}+(1/2)m\omega _{x}^{2}x^{2}+c_{0}^{(\mathrm{%
1D})}n_{\mathrm{tot}}$ is the spin-independent part of the Hamiltonian, with
$n_{\mathrm{tot}}=|\psi _{-1}|^{2}+|\psi _{0}|^{2}+|\psi _{+1}|^{2}$ being
the total density ($m$ is the atomic mass). The nonlinearity coefficients
have an effectively 1D form, namely $c_{0}^{(\mathrm{1D})}=c_{0}/2\pi
a_{\perp }^{2}$ and $c_{2}^{(\mathrm{1D})}=c_{2}/2\pi a_{\perp }^{2}$, where
$a_{\perp }=\sqrt{\hbar /m\omega _{\perp }}$ is the transverse
harmonic-oscillator length which determines the size of the transverse
ground state. Coupling constants $c_{0}$ and $c_{2}$ account for,
respectively, the mean-field spin-independent and spin-dependent binary
interactions between identical spin-$1$ bosons,
\begin{equation}
\left\{ c_{0},c_{2}\right\} =\left( 4\pi \hbar ^{2}/3m\right) \left\{
(a_{2}+a_{0}),(a_{2}-a_{0})\right\} ,  \label{c0c2}
\end{equation}%
where $a_{0}$ and $a_{2}$ are the $s$-wave scattering lengths in combined
symmetric collision channels of total spin $f=0,2$. The spinor $F=1$
condensate with $c_{2}<0$ and $c_{2}>0$ is, respectively, of the FM and AFM
types, as, respectively, in $^{87}$Rb and $^{23}$Na \cite{ho,ohmi}.

Measuring time, length and density in units of $\hbar /c_{0}^{(1D)}n_{0}$, $%
\hbar /\sqrt{mc_{0}^{(1D)}n_{0}}$ and $n_{0}$ (here, $n_{0}$ is the peak
density), we cast Eqs.~(\ref{mgp1})-(\ref{mgp2}) in the dimensionless form,
\begin{eqnarray}
i\partial _{t}\psi _{\pm 1} &=&H_{\mathrm{si}}\psi _{\pm 1}+\delta (|\psi
_{\pm 1}|^{2}+|\psi _{0}|^{2}-|\psi _{\mp 1}|^{2})\psi _{\pm 1}  \notag \\
&+&\delta \psi _{0}^{2}\psi _{\mp 1}^{\star },  \label{dvgp1} \\
i\partial _{t}\psi _{0} &=&H_{\mathrm{si}}\psi_{0}+\delta (|\psi
_{-1}|^{2}+|\psi _{+1}|^{2})\psi _{0}  \notag \\
&+&2\delta \psi _{-1}\psi _{0}^{\star }\psi _{+1},  \label{dvgp2}
\end{eqnarray}
where $H_{\mathrm{si}}\equiv -(1/2)\partial _{x}^{2}+(1/2)\Omega _{\mathrm{tr%
}}^{2}x^{2}+n_{\mathrm{tot}}$, the normalized trap's strength is
\begin{equation}
\Omega _{\mathrm{tr}}=\frac{3}{2(a_{0}+2a_{2})n_{0}}\left( \frac{\omega _{x}%
}{\omega _{\perp }}\right) ,  \label{Omegatr}
\end{equation}
and the parameter $\delta$ is given by 
\begin{equation}
\delta \equiv \frac{c_{2}^{(1D)}}{c_{0}^{(1D)}}=\frac{a_{2}-a_{0}}{%
a_{0}+2a_{2}}  \label{delta}
\end{equation}
($\delta <0$ and $\delta >0$ correspond, respectively, to the FM and AFM
spinor condensate).   
For spin-$1$ $^{87}$Rb and $^{23}$Na atoms, this parameter is
$\delta =-4.66\times 10^{-3}$ \cite{kemp} and $\delta =3.14\times
10^{-2}$ \cite{greene}, respectively. Thus, in these physically
relevant cases, $\delta $ is a fixed small parameter of the
system.

\textit{Spin-polarized states} of the system, characterized by a constant
population of each spin component, can be found upon substitution
\begin{equation}
\psi _{j}=\sqrt{n_{j}(x)}\exp \left( i\theta _{j}-i\mu _{j}t\right)
,\,\,\,j=-1,0,+1,  \label{spsa}
\end{equation}%
where $n_{j}$ and $\theta _{j}$ are densities and phases of the components,
and $\mu _{j}$ are their chemical potentials. Substituting this in Eqs.~(\ref%
{dvgp1}) and (\ref{dvgp2}), it is readily found that conditions for the
existence of spin-polarized states are
\begin{eqnarray}
2\mu _{0} &=&\mu _{-1}+\mu _{+1},  \\
\Delta \theta &\equiv &2\theta _{0}-\theta _{+1}-\theta _{-1}=0\,\,\,\,%
\mathrm{or}\,\,\,\pi .  \label{cond2}
\end{eqnarray}%
Below, we assume that the chemical potentials of the components are equal, $%
\mu \equiv \mu _{-1}=\mu _{0}=\mu _{+1}$. For analysis of the stability of a
stationary spin-polarized state, 
$\Psi _{\mathrm{sps}}(x)= [\tilde\psi_{-1}(x),\tilde\psi_{0}(x),\tilde\psi _{+1}(x)]^{T}$, 
we will perform
linearization around the unperturbed spin-polarized state, assuming a
perturbed solution,
\begin{equation}
\psi _{j}(x,t)=\left[ \tilde\psi_j(x)+\epsilon \left(
u_{j}(x)e^{-\lambda _{j}t}+\upsilon _{j}^{\ast }(x)e^{\lambda _{j}^{\star
}t}\right) \right] e^{-i\mu t},  \label{stansatz}
\end{equation}%
where $u_{j}$ and $v_{j}$ represent infinitesimal perturbations with
eigenvalues $\lambda \equiv \lambda _{r}+i\lambda _{i}$. Then, the solution of
the ensuing linear-stability problem (i.e., the BdG equations) for $\lambda $
and associated eigenfunctions $u_{j},v_{j}^{\ast }$ provides complete
information about the stability of the underlying stationary state, $\Psi _{%
\mathrm{sps}}$. Whenever it is unstable, we will also examine its evolution
through direct simulations of GP equations (\ref{dvgp1}) and (\ref{dvgp2}),
using a finite difference scheme in space and a 
fourth-order Runge-Kutta time integrator. 
Typically, in the simulations the unstable state is initially perturbed by a
uniformly distributed random perturbation of relative amplitude $5\times
10^{-4}$.

To estimate relevant physical parameters, we assume the spinor
condensate of $^{23}$Na atoms with peak 1D density $n_{0}=10^{8}$ m$^{-1}$,
confined in the harmonic trap with frequencies $\omega _{\perp }=2\pi \times
230~$Hz and $\omega _{x}=2\pi \times 13~$Hz. In this case, the normalized
trap strength is $\Omega =0.1$ (this value is kept fixed throughout this
work), while the number of atoms, $N$, depends on chemical potential $\mu $
and the particular form of spin-polarized states. Typically, we take $\mu $
in interval $1\leq \mu \leq 5$, which corresponds to values of $N$ in the
range of $3.5\times 10^{3}\leq N\leq 3.5\times 10^{4}$ atoms. For instance, at $%
\mu =2$ the number of atoms is $N\sim 10^{4}$; in this case, the normalized
time and space units correspond, respectively, to $1.2$ ms and $1.83~\mu $m.

\section{Thomas-Fermi spin-polarized states}

The simplest spin-polarized states can be found in the framework of the
\textit{single-mode approximation} \cite{sm,ofy2}. In anticipation,
suggested by the approximation that components $\sqrt{n_{j}(x)}$ may be
close to eigenmodes of a single effective potential, induced by the
combination of the trap and nonlinearity, we introduce the ansatz $%
n_{j}(x)=q_{j}n(x)$, where coefficients $q_{j}$ are the populations of each
spin component in the steady state, related by the normalization condition, $%
q_{-1}+q_{0}+q_{+1}=1$. Then, Eqs.~(\ref{dvgp1})-(\ref{dvgp2}) lead to the
following system:
\begin{eqnarray}
\left[ \hat{L}+\delta \left( 1+\sqrt{\frac{q_{\mp 1}}{q_{\pm 1}}}(sq_{0}-2%
\sqrt{q_{-1}q_{+1}})\right) n\right] \sqrt{n}=0,\quad  \label{in1}
\\[2.0ex]
\left[ \hat{L}+\delta \left( 1-q_{0}+2s\sqrt{q_{-1}q_{+1}}\right) n\right]
\sqrt{n}=0, \quad \label{in2}
\end{eqnarray}
where $\hat{L}\equiv -\mu -(1/2)\partial _{x}^{2}+(1/2)\Omega ^{2}x^{2}+n$
and $s=\pm 1$ for $\Delta \theta =0$ or $\Delta \theta =\pi $, respectively.
It is readily observed that the three equations (\ref{in1}), (\ref{in2}) are
tantamount to a single one, viz. 
\begin{equation}
\left( -\mu -\frac{1}{2}\partial _{x}^{2}+\frac{1}{2}\Omega
^{2}x^{2}+pn\right) \sqrt{n}=0,  \label{gs}
\end{equation}%
supplemented by relations
\begin{eqnarray}
&&p\equiv 1+\delta ,\,\,\,\ q_{0}=2\sqrt{q_{-1}q_{+1}}\,\,\,\ \mathrm{for}%
\,\,\,\ \Delta \theta =0,  \label{con1} \\[1.0ex]
&&p\equiv 1,\,\,\,\ q_{-1}=q_{+1}\,\,\,\ \mathrm{for}\,\,\,\ \Delta \theta
=\pi .  \label{con2}
\end{eqnarray}%
%
%
%

In uniform space ($\Omega =0$), Eq.~(\ref{gs}) leads to the constant
density, $n=\mu /p$. As shown in Refs. \cite{ofy2,boris,ourspin}, such
constant solutions to Eqs.~(\ref{dvgp1})-(\ref{dvgp2}) are modulationally
stable only in the AFM phase, with $\delta >0$. Below, we will only consider
the case of the AFM condensate (in particular, bosonic sodium atoms with
spin $1$ and $\delta =3.14\times 10^{-2}$). In the presence of a sufficiently
weak trap, Eq.~(\ref{gs}) can be solved approximately employing the TF
approximation \cite{pit}: neglecting the kinetic-energy term ($\sim \partial
_{x}^{2}\sqrt{n}$) in Eq.~(\ref{gs}), we find density profiles of the three
spin components: in the region where $\mu >(1/2)\Omega ^{2}x^{2}$,
\begin{equation}
n_{j}=\frac{q_{j}}{p}\left( \mu -\frac{1}{2}\Omega ^{2}x^{2}\right) ,
\label{TF}
\end{equation}%
and $n_{j}=0$ elsewhere. Obviously, all three components of the TF
solution have the same spatial width, or Thomas-Fermi radius, 
$R_{\mathrm{TF}}=\sqrt{2\mu }/\Omega $.

\begin{figure}[tbp]
\includegraphics[width=4.12cm,height=3.3cm]{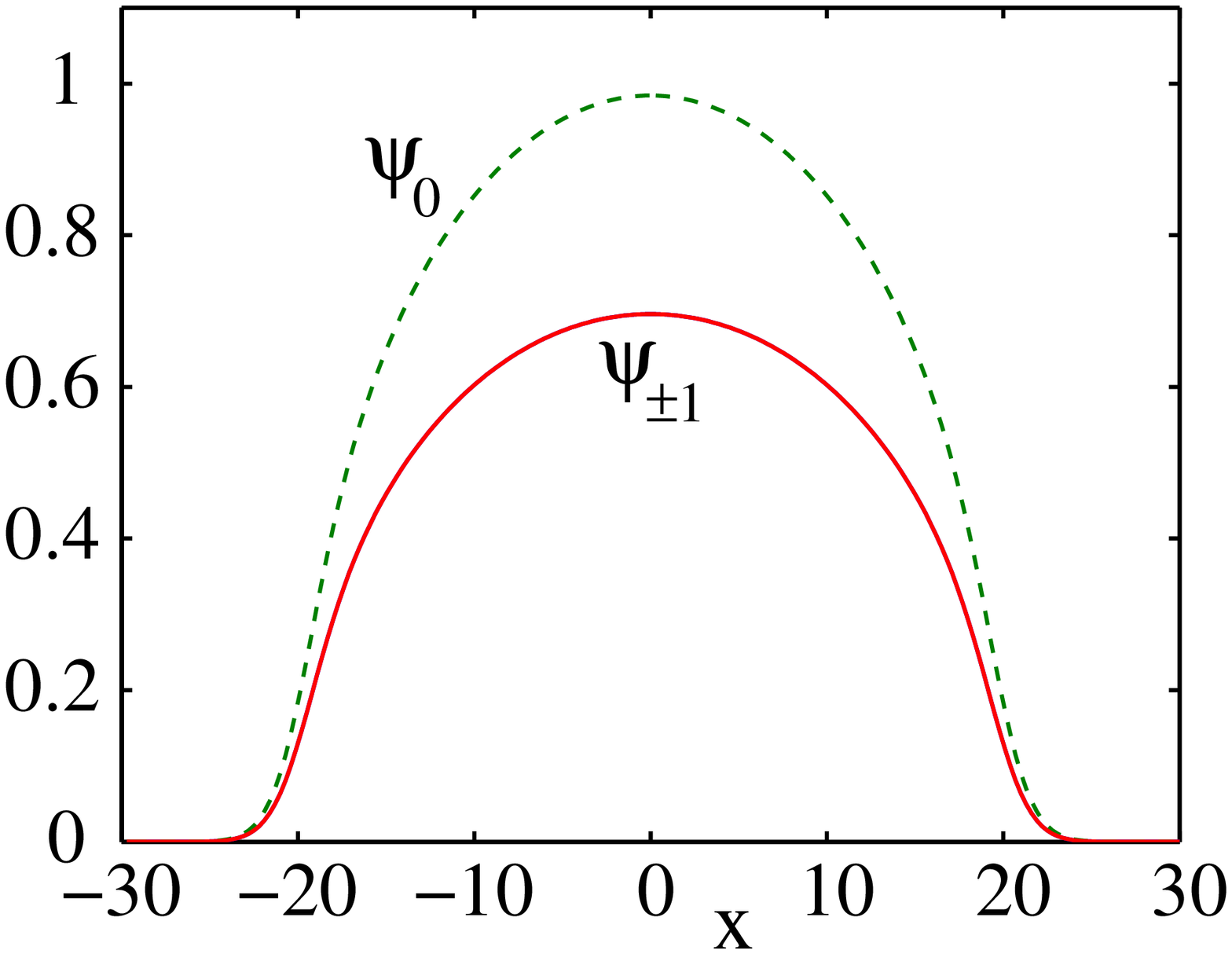}
~~
\includegraphics[width=3.9cm,height=3.4cm]{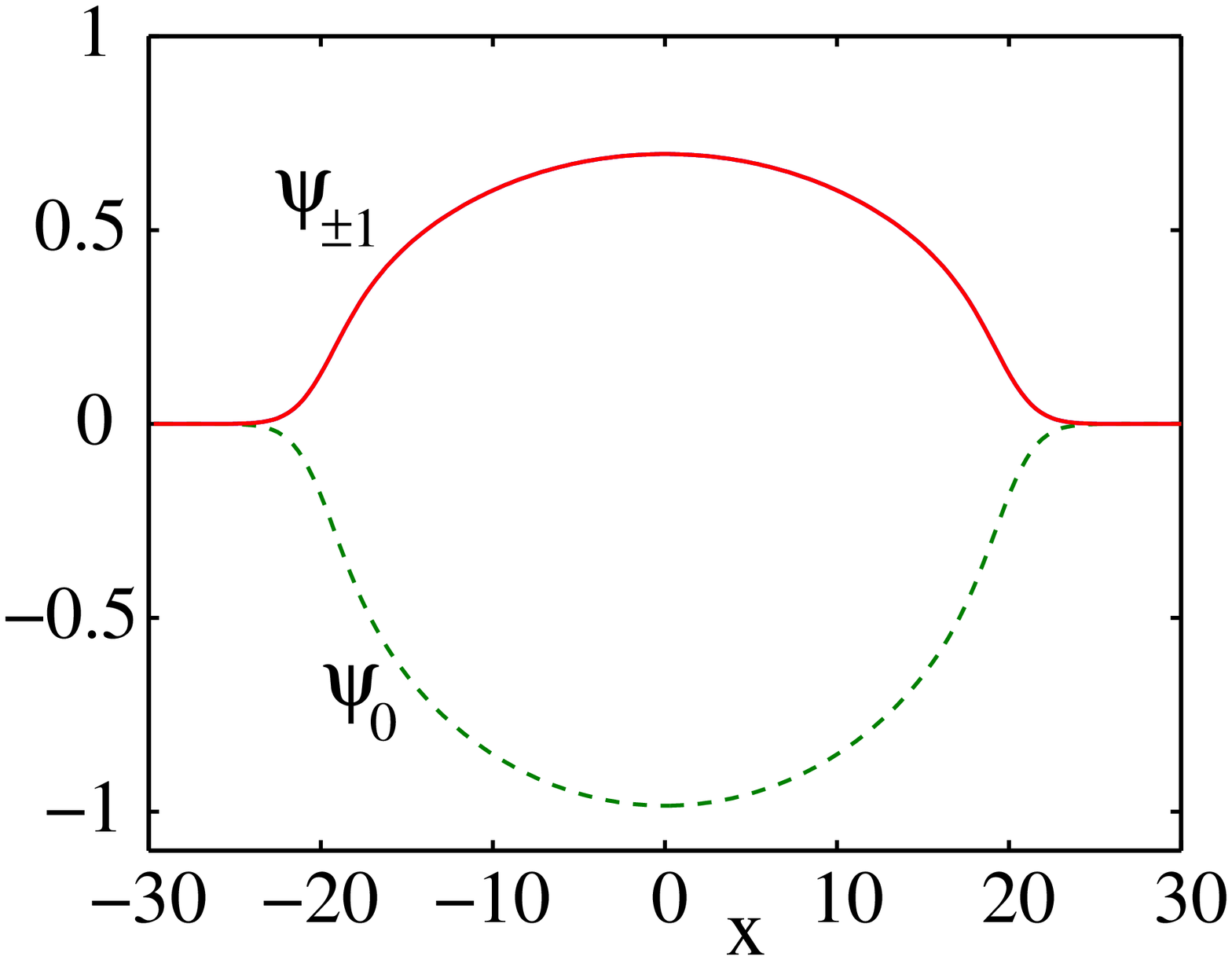}
\caption{(Color online) Examples of two stable spin-polarized
states for $\Delta \protect\theta =0$ (left panel) and $\Delta \protect%
\theta =\protect\pi $ (right panel), both obtained for $\Omega =0.1$ and $%
\protect\mu =2$. The wavefunctions $\protect\psi _{\pm 1}$ are identical and
are depicted by the solid line, while the wavefunction $\protect\psi _{0}$
is depicted by the dashed line. In the left panel, $q_{-1}=q_{+1}=0.5$ and $%
q_{0}=1$, while in the right panel $q_{-1}=q_{+1}=0.25$ and $q_{0}=0.5$.
}
\label{fig1}
\end{figure}

In numerical simulations, we used a fixed-point algorithm (Newton-Raphson method)
to find exact spin-polarized solutions to Eqs.~(\ref{dvgp1})-(\ref{dvgp2}),
with profiles close to those produced by the TF approximation, as given by
Eq.~(\ref{TF}). In particular, we used, as an initial guess, three identical
profiles of the form
\begin{equation}
\psi _{j}(x)=\sqrt{n(x)}\exp (i\theta _{j}),  \label{sa}
\end{equation}
with $n(x)=\mu -(1/2)\Omega ^{2}x^{2}$ and $\Delta \theta =0$ or $\pi $.
Then, keeping the trap's strength, $\Omega $, fixed, we varied the chemical
potential $\mu $, and the numerical solution converged to stable spin-polarized
states, which were indeed close to the approximate one given by Eq.~(\ref{TF}%
). Two typical examples are shown in the top panels of Fig.~\ref{fig1} for
both cases, $\Delta \theta =0$ (left panel) and $\Delta \theta =\pi $ (right
panel), with $\Omega =0.1$ and $\mu =2$. The numerically determined states are
very close to their TF-predicted counterparts,
with $q_{-1}=q_{+1}=0.5$ and $%
q_{0}=1$ ($\Delta \theta =0$, left panel) and $q_{-1}=q_{+1}=0.25$ and $%
q_{0}=0.5$ ($\Delta \theta =\pi $, right panel).

\section{Spin-polarized states with holes}

Apart from the TF spin-polarized states, there exist other ones, which
feature holes in some of the components. 
%
An example of such states is shown in Fig.~\ref{fig2}. 
As seen in the two top panels of this figure (both a stable and an unstable
state are shown ---see below), the $\psi _{0}$ component is concentrated in
the form of a pulse located at the trap's center, while the $\psi _{\pm 1}$
components feature a large hole at the same spot. This arrangement \ is
explained by the fact that the interaction between components is repulsive,
hence a peak (hole) in $\psi _{0}$ ($\psi _{\pm 1}$) induces a hole (peak)
in $\psi _{\pm 1}$ ($\psi _{0}$). The norm $N$ of each component is shown,
as a function of chemical potential $\mu $, in the third-row panel of Fig.~%
\ref{fig2}. The set of the linear stability eigenvalues for this state is shown in
the second-row panels of Fig.~\ref{fig2}, for two different values of $\mu $%
: The state with $\mu =3$ (top-right) is stable, as all the eigenvalues are
imaginary, while the state with $\mu =2$ (top-left panel) is unstable. In fact, all such
unstable states are destabilized by a Hamiltonian Hopf bifurcation, which
leads to a quartet of eigenvalues with nonzero real parts. The instability
interval is $1.81\leq \mu \leq 2.15$, with the maximum instability growth
rate Max$\left( \lambda _{r}\right)\approx 1.3\times 10^{-3}$, found at
$\mu \approx 2$ (see bottom panel of Fig.~\ref{fig2}).

\begin{figure}[tbp]
~~\includegraphics[width=8cm]{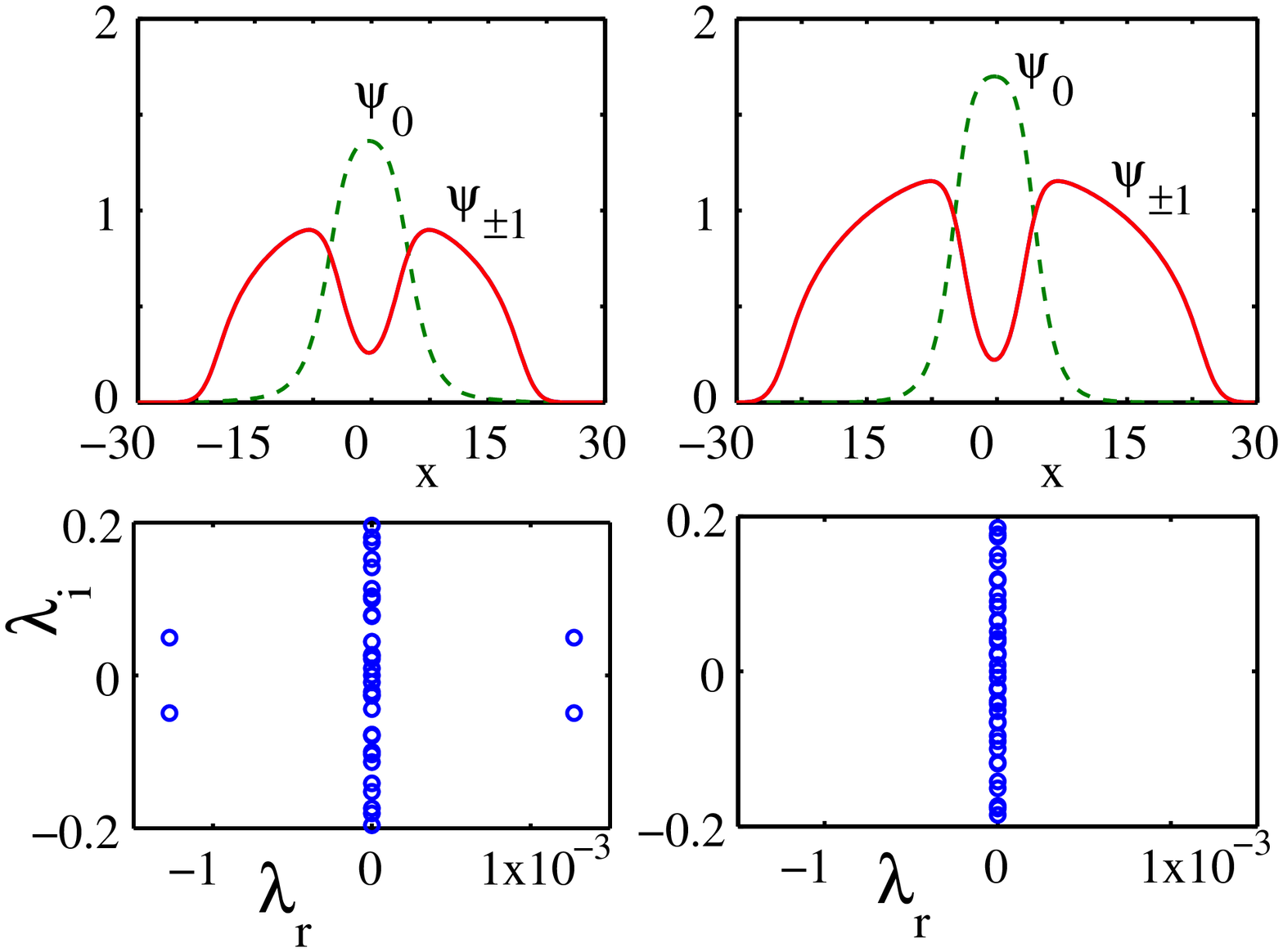} \newline
\includegraphics[width=8cm]{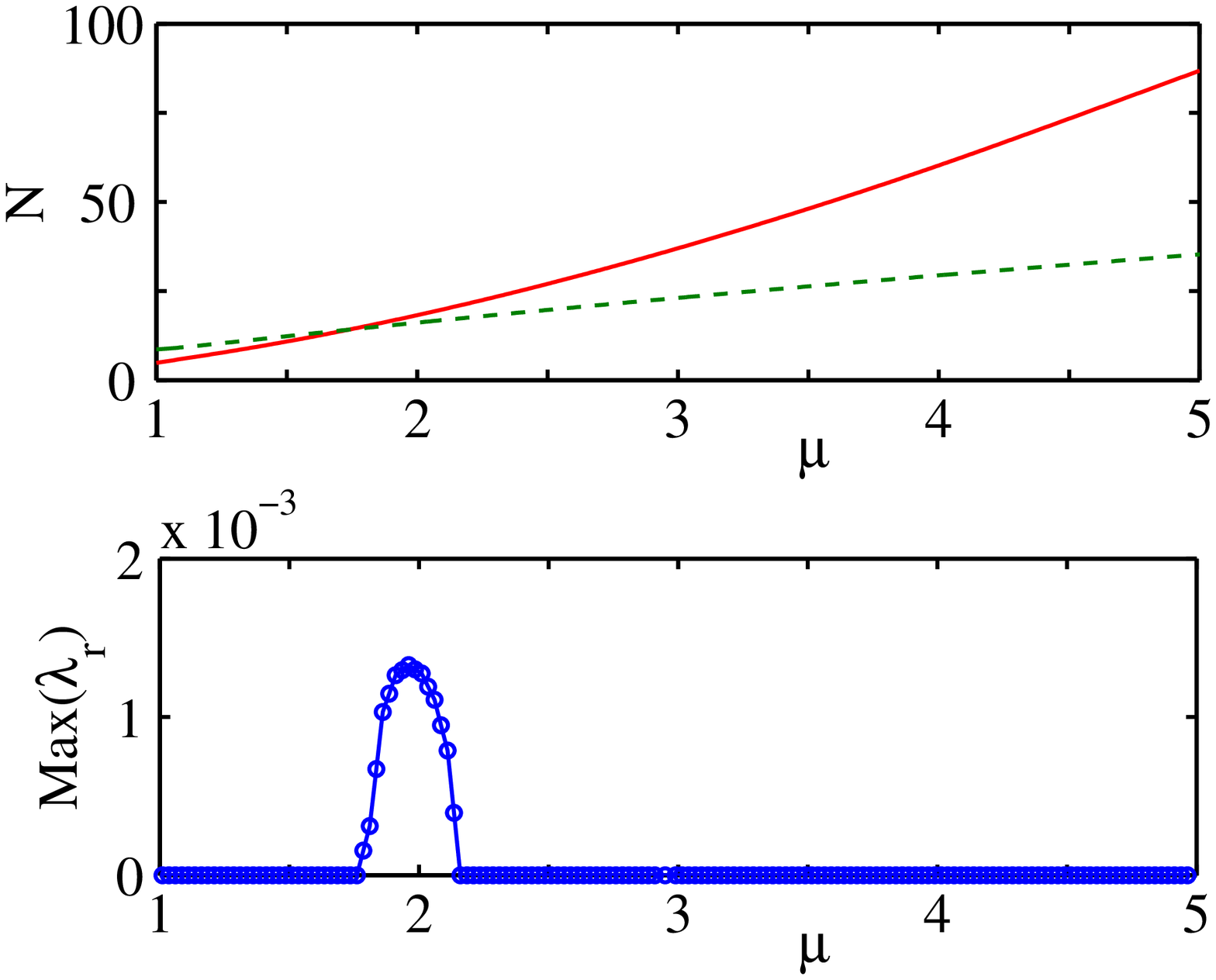} ~
\caption{(Color online) The spin-polarized state with a hole in each of the $%
\protect\psi _{\pm 1}$ components, and a pulse-like $\protect\psi _{0}$
component, for $\Omega =0.1$ and $\Delta \protect\theta =0$. The two top left
and right panels show unstable and stable states, with $\protect\mu =2$ and $%
\protect\mu =3$, respectively; solid and dashed lines depict the components $%
\protect\psi _{\pm 1}$ and $\protect\psi _{0}$. The two panels in the second row
display the spectral planes $\left( \protect\lambda _{r},\protect\lambda %
_{i}\right) $ of the (in)stability eigenvalues for the same states. Note
that the instability (at $\protect\mu =2$) is of the oscillatory type,
accounted for by a quartet of eigenvalues with nonzero real parts. The panel
in the third row shows the normalized number of atoms (norm), $N$, of each
component as a function of chemical potential $\protect\mu $; solid and
dashed lines show the norms of components $\protect\psi _{\pm 1}$ and $%
\protect\psi _{0}$, respectively. The bottom panel shows $\protect\lambda %
_{r}$ as a function of $\protect\mu $, which reveals the instability window
at $1.81\leq \protect\mu \leq 2.15$, with a maximum instability growth rate $
\left( \protect\lambda _{r}\right) _{\max }\approx 1.3\times 10^{-3}$ at $%
\protect\mu \approx 2$. The latter value corresponds to the unstable state
shown in the top-left panel.}
\label{fig2}
\end{figure}

A similar state 
with the roles of $\psi _{\pm 1}$ 
and $\psi _{0}$ exchanged, i.e., the $\psi _{0}$ component featuring the hole,
and $\psi _{\pm 1}$ ones concentrated in narrow pulses, were also found. Moreover, such states were found 
with either $\psi_0$ or $\psi_{\pm 1}$ having the opposite sign (i.e., for $\Delta \theta =\pi$). 
The results are not shown here, as the (in)stability of these states is 
qualitatively the same as in the above case. 
%

\begin{figure}[tbp]
\includegraphics[width=4.1cm]{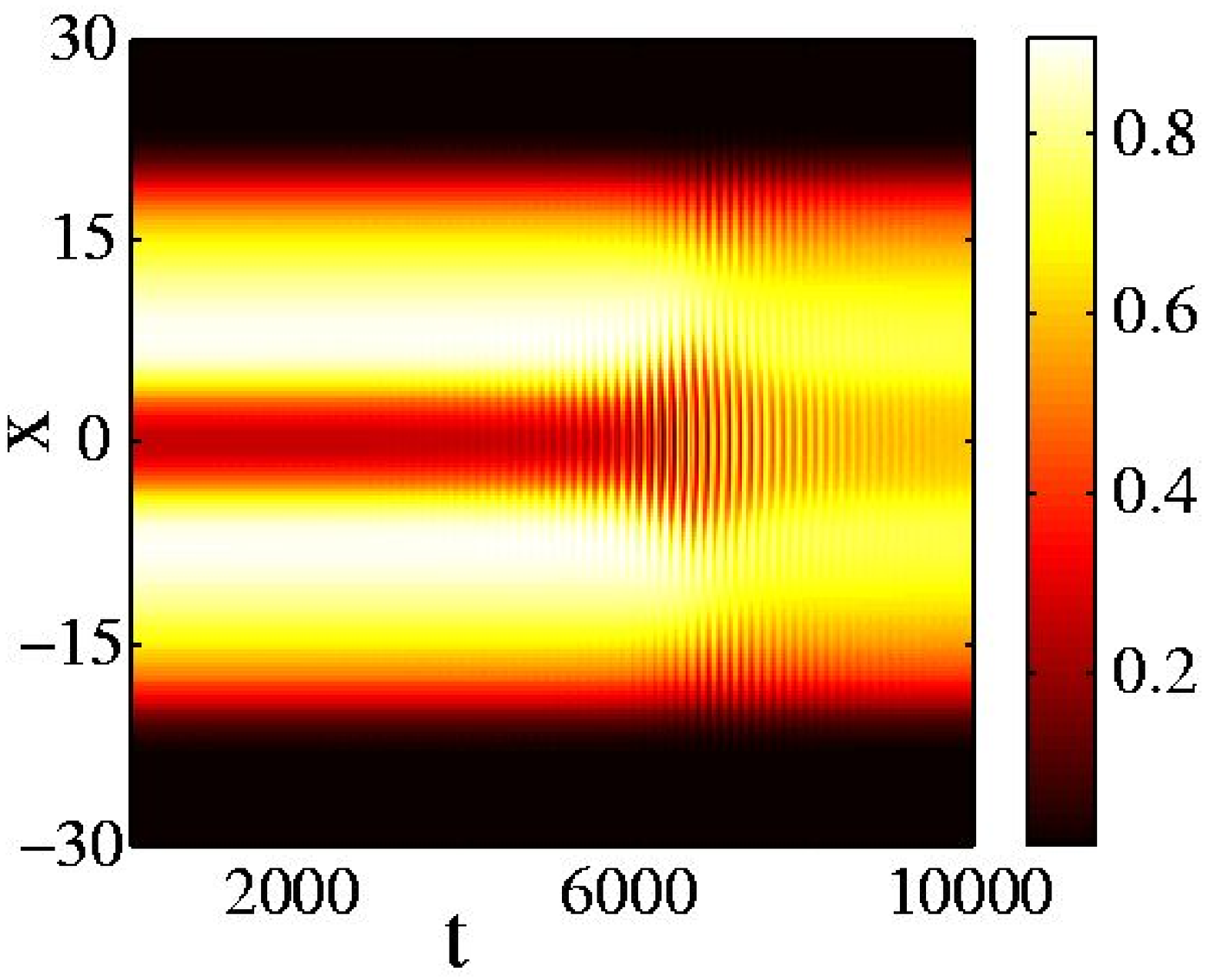} %
~
\includegraphics[width=4.1cm,height=3.4cm]{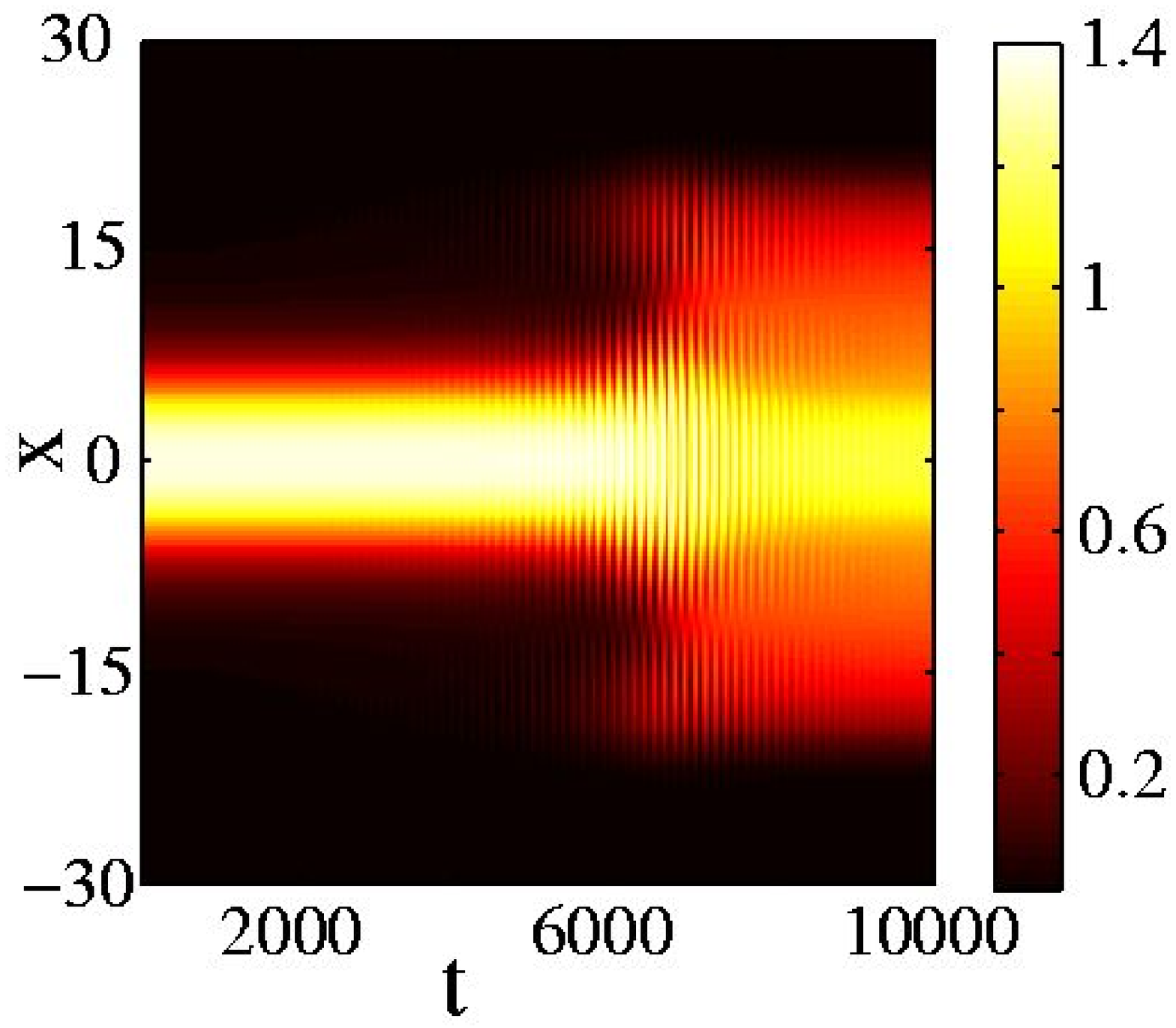}
\includegraphics[width=4.1cm]{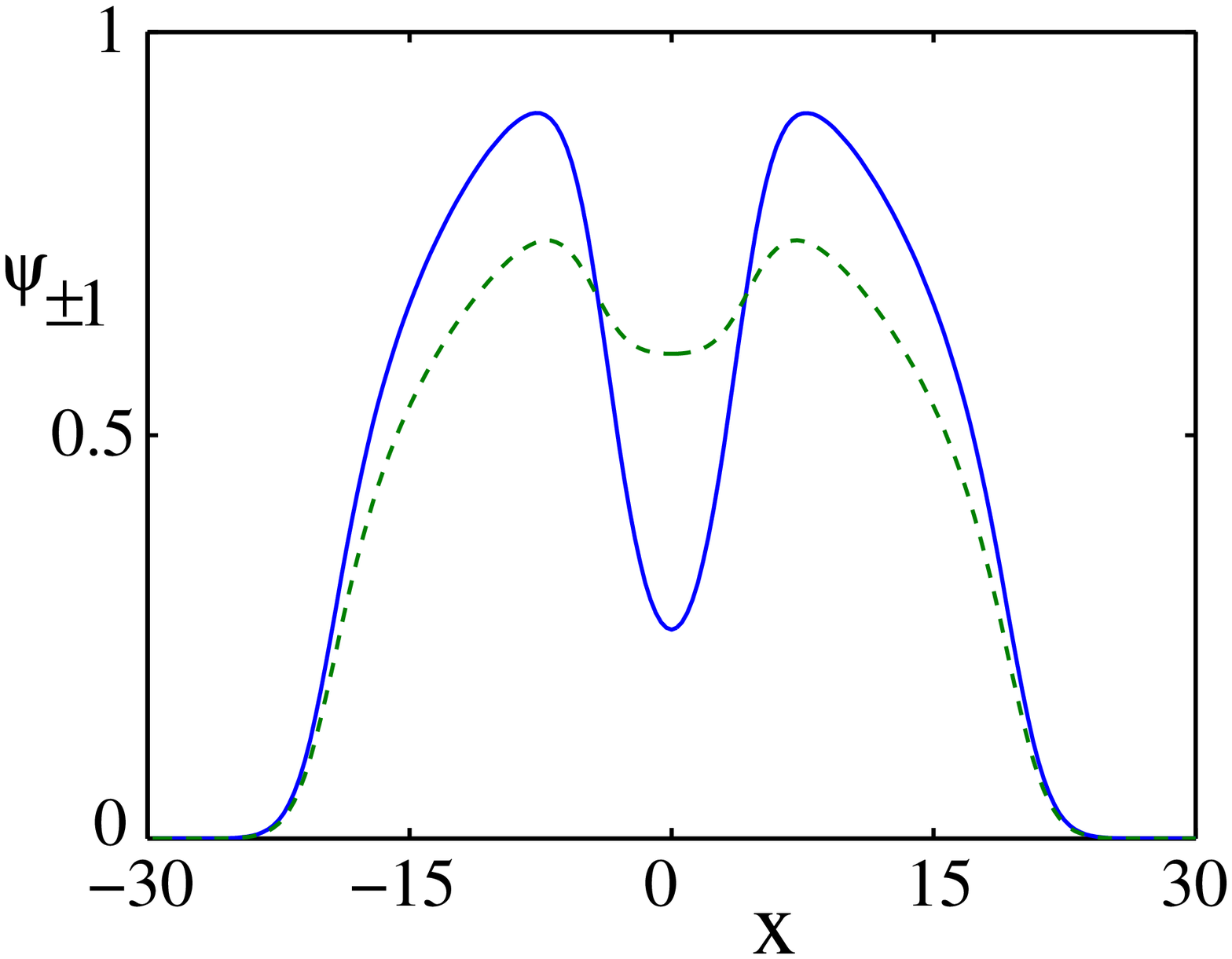}
~
\includegraphics[width=4.1cm]{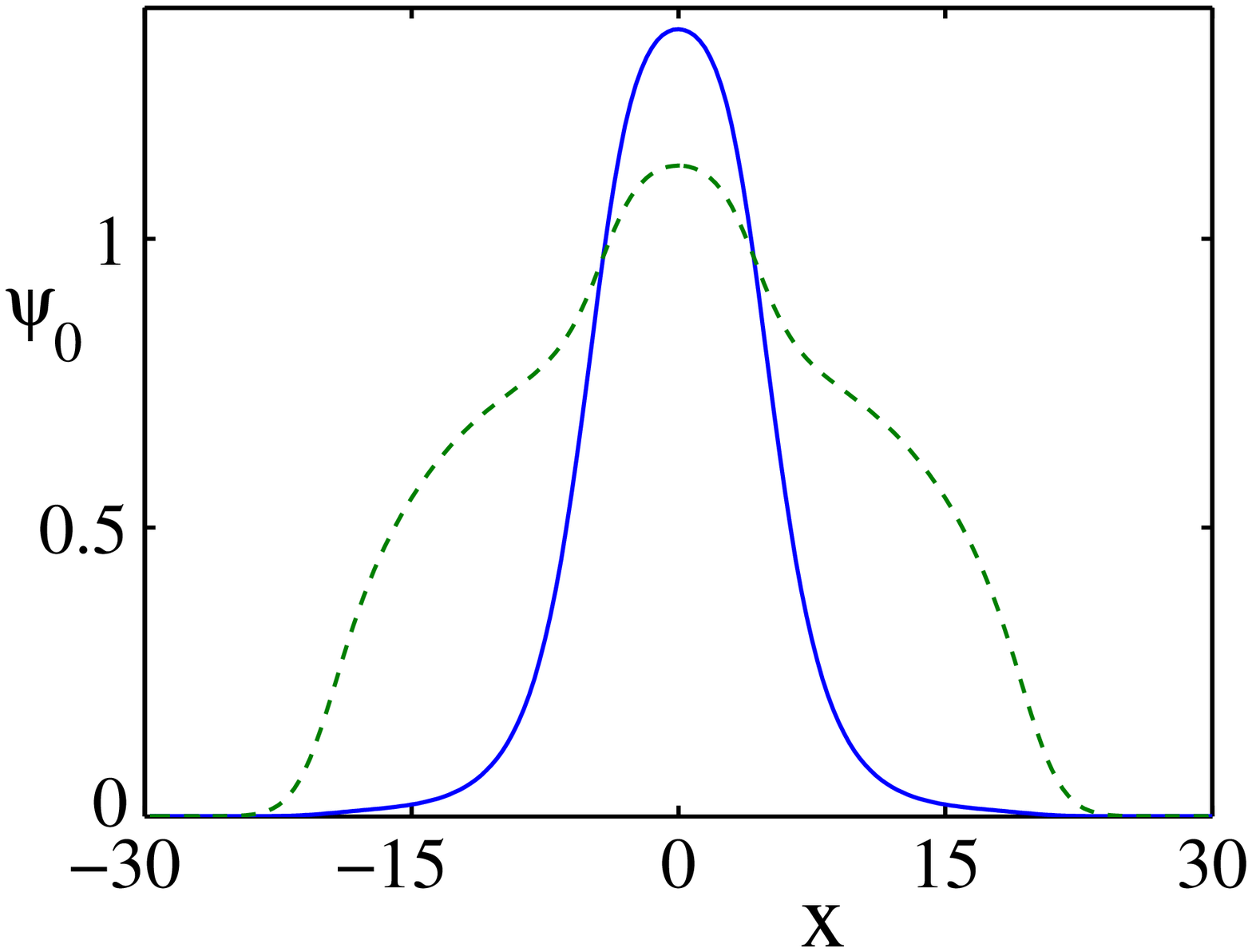}
\caption{(Color online) Top panels: Contour plots of densities of a solution shown in Fig.~\ref{fig2} 
subject to the oscillatory instability. The left and right panels
display, respectively, identical densities of $\protect\psi _{\pm 1}$ and
that of $\protect\psi _{0}$. Because of the extremely small growth rate of
the instability, it manifests itself only for $t>4000$. 
Bottom panels: The respective wavefunction profiles at $t=0$ (solid lines) and $t=10000$ (dashed lines);  
as above, the left and right panels show, respectively, the wavefunctions $\protect\psi _{\pm 1}$ 
and $\protect\psi _{0}$.  
}
\label{fig4}
\end{figure}

The evolution of unstable states is exemplified in Fig.~\ref{fig4}, which
displays results of direct simulations of Eqs.~(\ref{dvgp1}) and (\ref{dvgp2}%
) for the unstable state with $\Delta \theta =0$ and $\mu =2$, that was
presented in Fig.~\ref{fig2} (top-left panel). 
In Fig.~\ref{fig4}, contour plots of densities
of the components of the solution are displayed as a function of time (the
densities of the $\psi _{+1}$ and $\psi _{-1}$ components are identical). It
is clearly observed that the predicted oscillatory instability sets in at 
a very large time ($t\approx 4000$, which corresponds to $t\approx 5$
seconds in physical units); this is a consequence of the extremely small
growth rate of the instability. Eventually, the system settles down to a
steady state, which has many similarities with the initial one. In
particular, as seen in the top panels of Fig.~\ref{fig4}, after $t\approx 8000$ the pulse in
the $\psi _{0}$ component broadens and its amplitude is accordingly
decreased, while the hole in the $\psi _{\pm 1}$ components becomes
shallower. It is also noted that the three components develop a similar
structure in their tails, as shown in the bottom panels of Fig.~\ref{fig4}.

Apart from the states considered above, it is also possible to find
spin-polarized states which feature, e.g., one hole in each of the $\psi
_{\pm 1}$ components, and two holes in $\psi_{0}$. Examples of such a state
are shown in the top panels of Fig.~\ref{fig5} (the left one, for $\mu =2$, is
stable, while the right one, for $\mu =3$ is unstable ---see below). As seen
in this figure, one may consider $\psi _{-1}$ and $\psi _{+1}$ as built of
two overlapping pulses, which induce two holes in the $\psi _{0}$ component
due to the repulsive interatomic interactions. Results of the stability
analysis for these states are shown in Fig.~\ref{fig5}. In this case, two 
quartets of eigenvalues with nonzero real parts are found in the spectral plane
(see the second row right panel in Fig.~\ref{fig5}) in the interval $2.58\leq
\mu \leq 3.22$ (the bottom panel in Fig.~\ref{fig5}). The respective largest 
instability growth rate is $\left( \lambda _{r}\right) _{\max }\approx
1.8\times 10^{-3}$ for $\mu \approx 2.9$, i.e., of the same order of
magnitude as in the previous case. The development of the instability was
studied, as above, in direct simulations, starting with\ initial conditions
in the form of a perturbed solution pertaining to $\mu =3$. The result is
shown in Fig.~\ref{fig6}, in terms of the evolution of identical densities
of the $\psi _{\pm 1}$ components, and the density of $\psi _{0}$. Again,
the instability manifests itself at large times ($t>3500$, which corresponds
to $t>4.2$ seconds in physical units), but in this case the final result is
a strong oscillatory deformation of the three spin components (after $t\approx 5500$), 
contrary to what is the case in Fig.~\ref{fig4} (the establishment of a
new spin-separated state).

\begin{figure}[tbp]
~ \includegraphics[width=8cm]{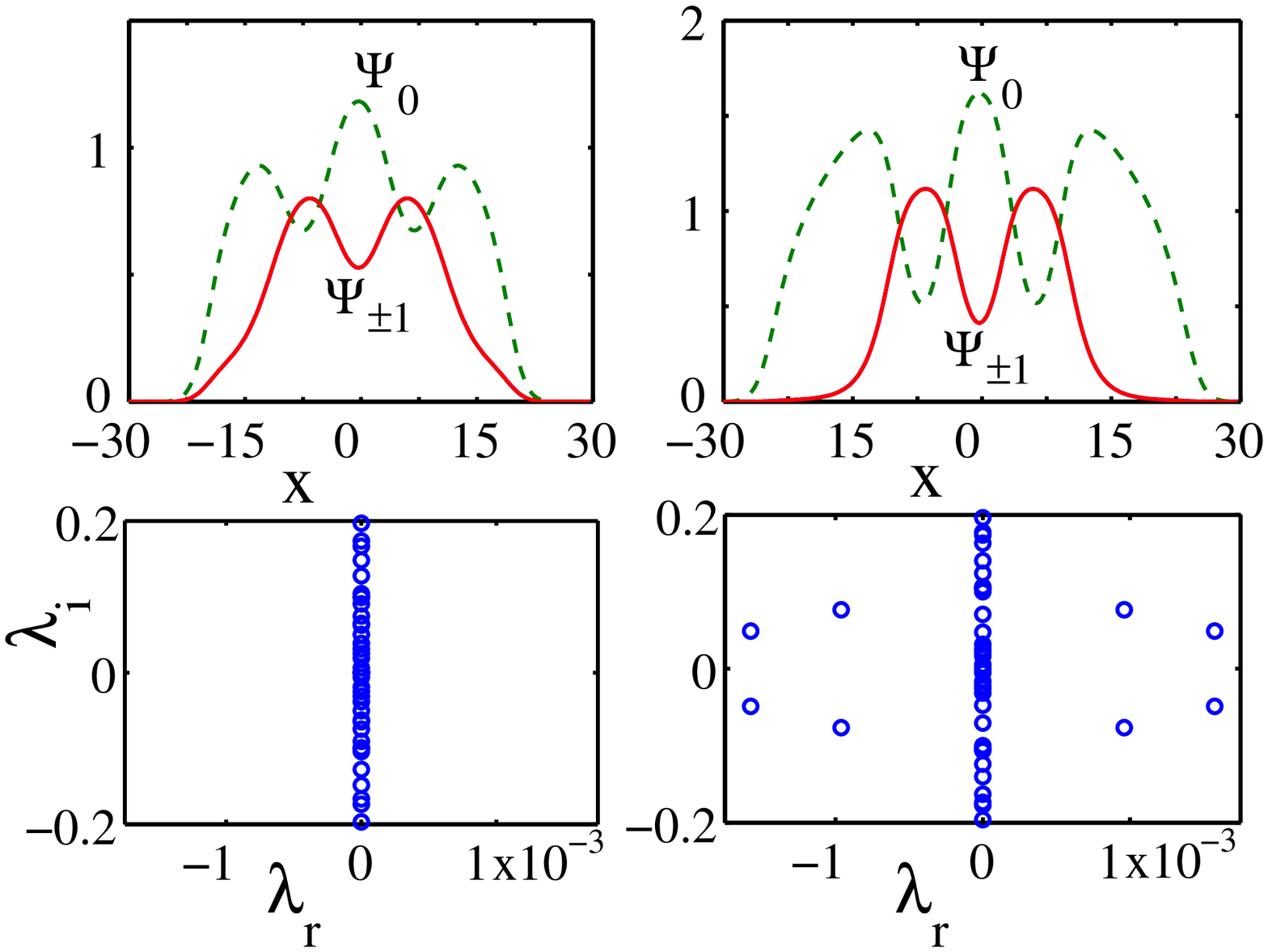} \newline
\includegraphics[width=8cm]{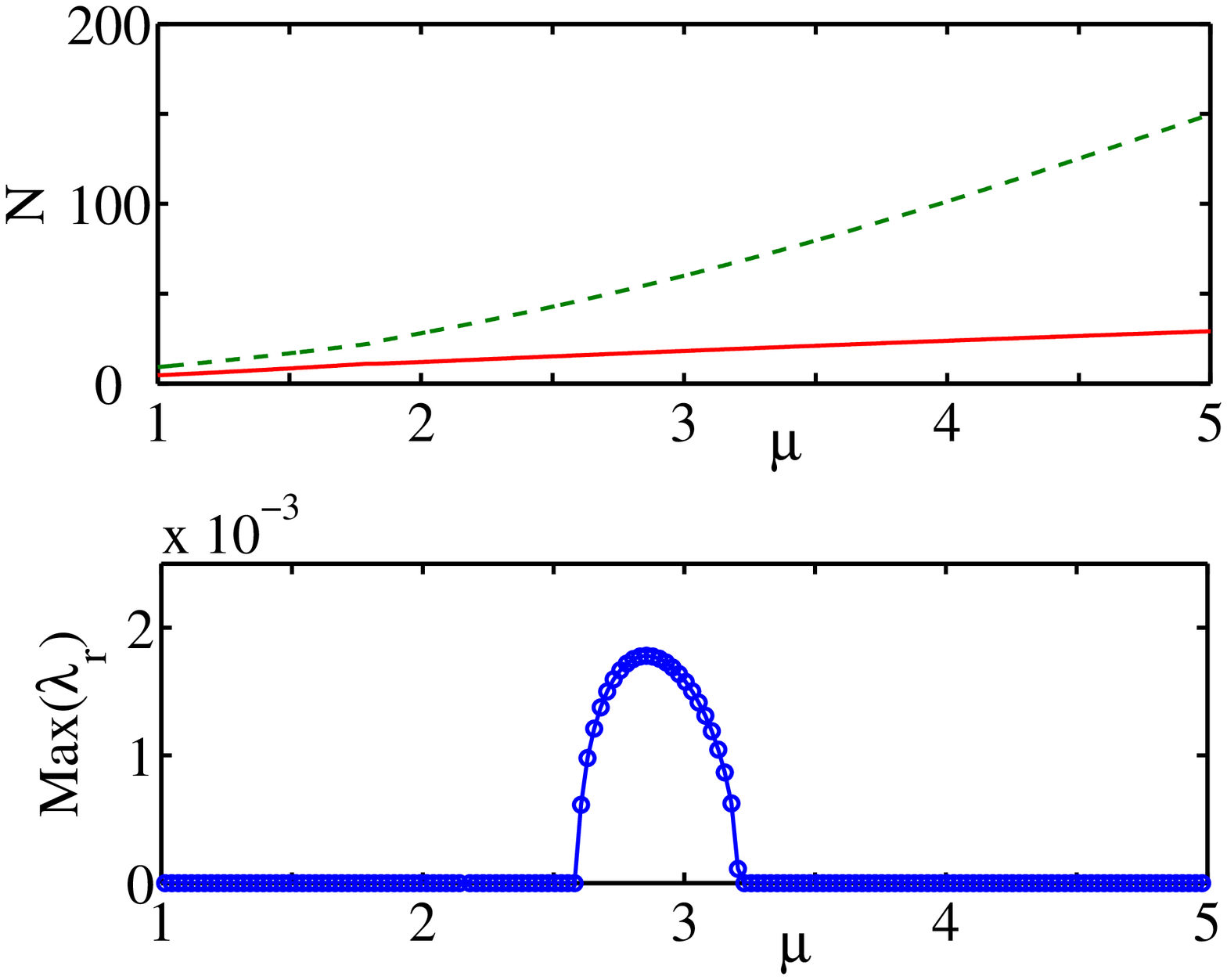}
\caption{(Color online) Same as Fig.~\protect\ref{fig2} but for a state with
one hole in each of the $\protect\psi _{\pm 1}$ components and two holes in
the $\protect\psi _{0}$ component. In this case, the instability is manifested
by two eigenvalue quartets the largest of which has a 
nonzero real part in the interval $2.58\leq \protect\mu \leq 3.22$;
the maximum instability growth rate is $\left( \protect\lambda _{r}\right)
_{\max }\approx 1.8\times 10^{-3}$ for $\protect\mu \approx 2.9$. The
unstable state shown in the top-right panel corresponds to $\protect\mu =3$.
}
\label{fig5}
\end{figure}

\begin{figure}[tbp]
\includegraphics[width=4.1cm,height=3.4cm]{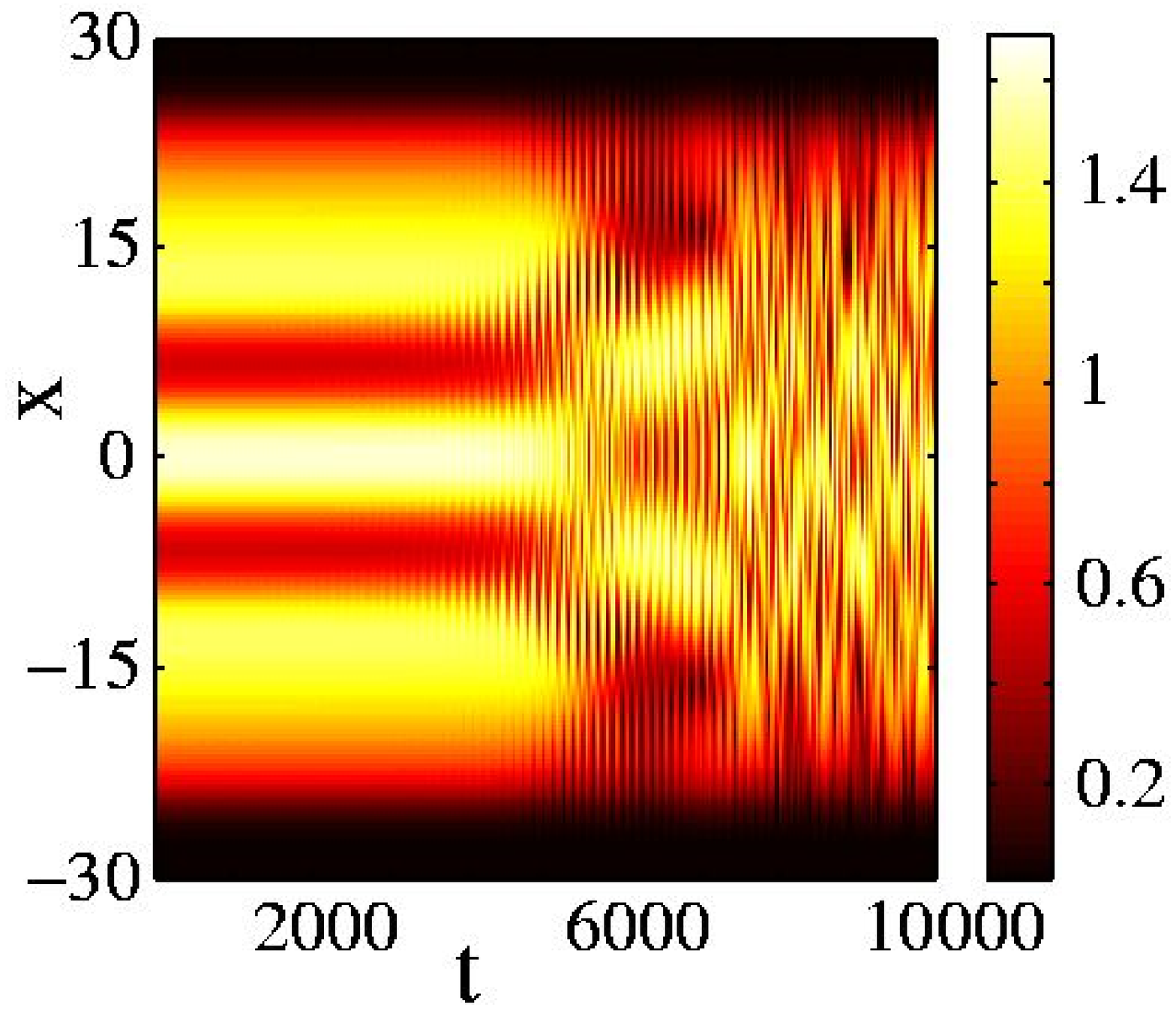}
~
\includegraphics[width=4.1cm,height=3.4cm]{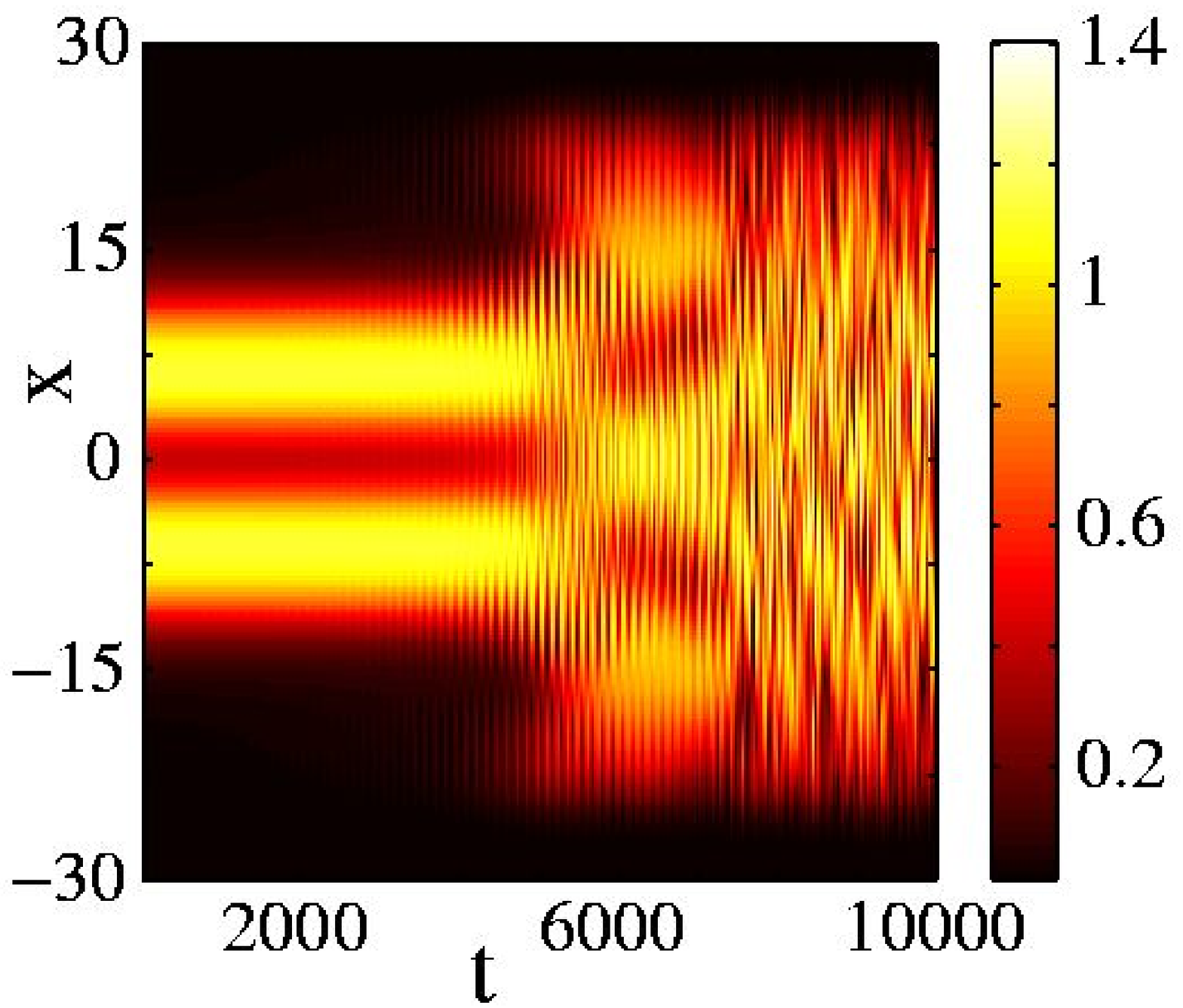}
\includegraphics[width=4.1cm]{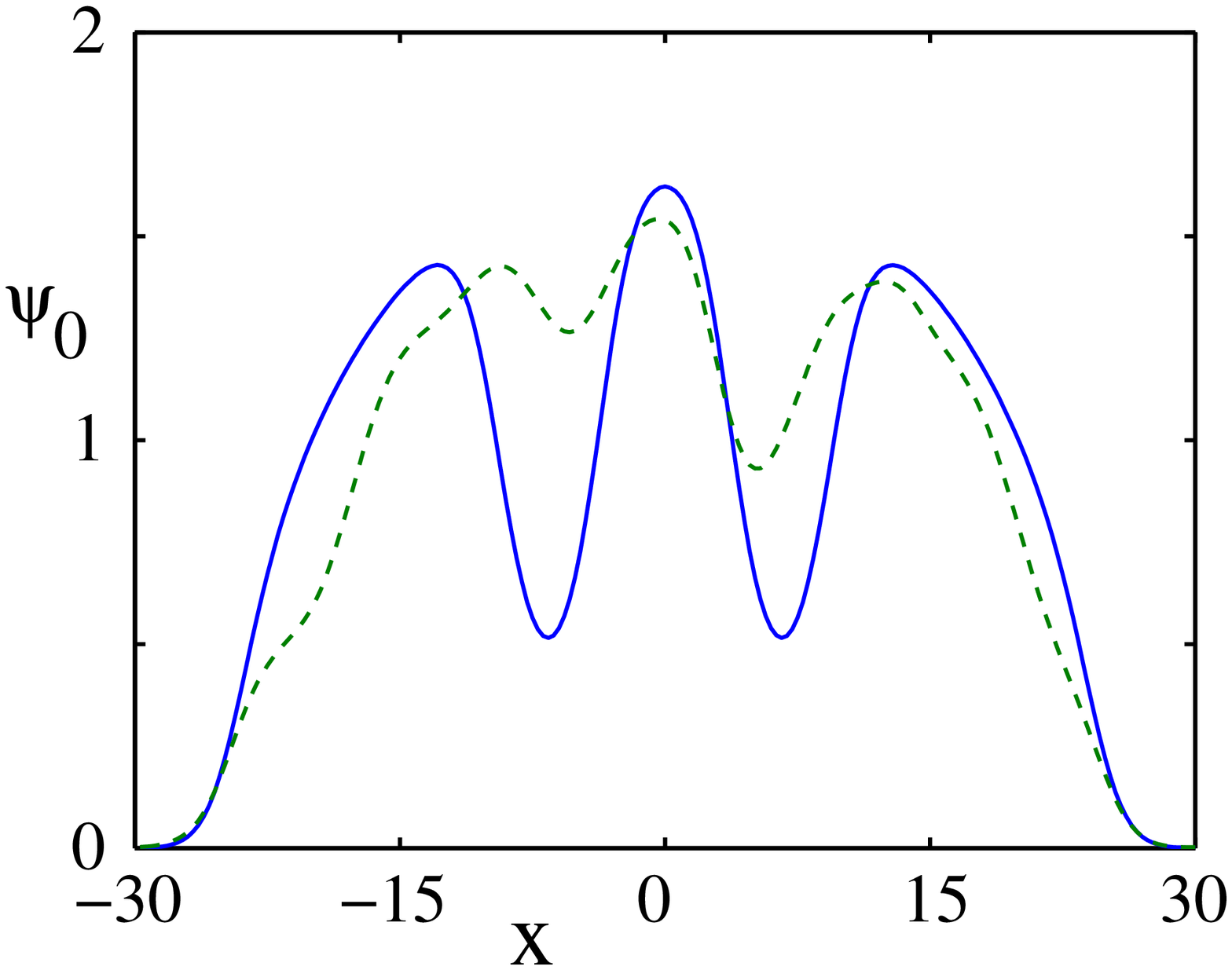}
~
\includegraphics[width=4.1cm]{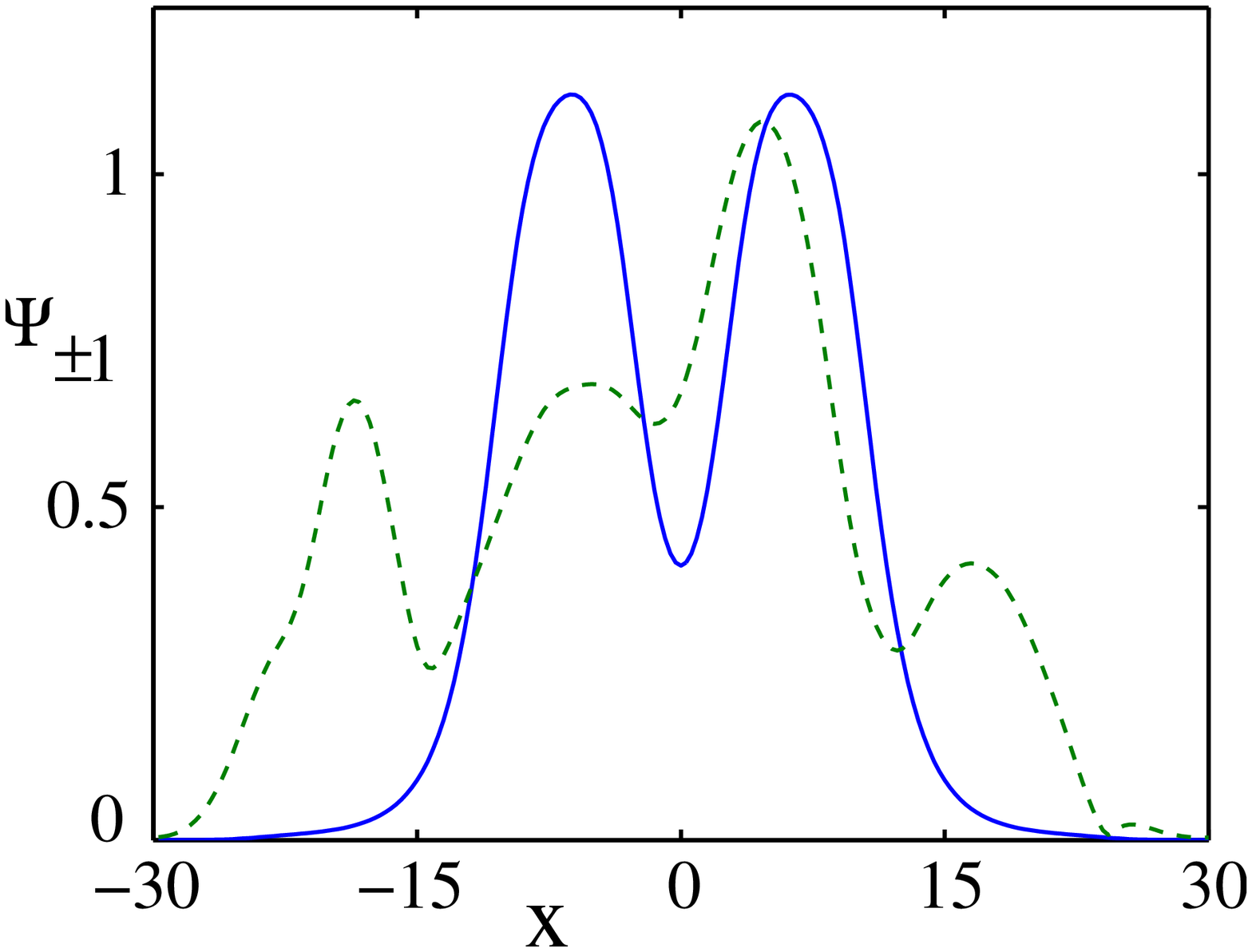}
\caption{(Color online) Same as Fig.~\protect\ref{fig4}, for the unstable
state shown in the top-right panel of Fig.~\protect\ref{fig5} (pertaining 
to $\protect\mu =3$). The instability manifests itself at large times ($%
t>3500$) and results in strong oscillatory deformation of the clouds. }
\label{fig6}
\end{figure}

Similar states with one hole in $\psi _{0}$ and two holes in each of the $%
\psi _{\pm 1}$ components, as well their counterparts with $\Delta \theta
=\pi $, have also been found. They are not shown here; as in the previous
case, the (in)stability of these additional states is similar to that
reported in Fig.~\ref{fig5}.

\begin{figure}[tbp]
\includegraphics[width=6.0cm]{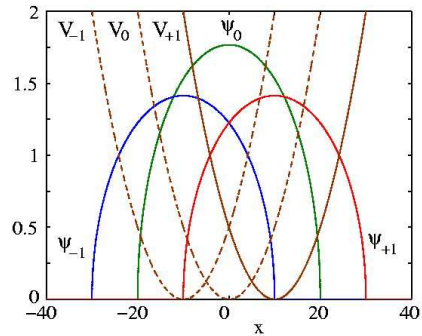}
\caption{(Color online) Initialization of the system when three different
traps of the same strength but of different location are used: the TF clouds
for $\protect\psi _{-1}$, $\protect\psi _{0}$ and $\protect\psi _{+1}$ 
are confined in
harmonic traps of strength $\Omega =0.1$ located, respectively, at $x=-10$, $%
x=0$ (dashed lines) and $x=10$ (solid line). The traps at $x=-10$ and $x=0$
are turned off and this spinor configuration, confined solely in the trap
located at $x=10$, is introduced in the fixed-point algorithm. }
\label{fig7}
\end{figure}

\section{Domain Walls}

In the above sections, we reported the spin-polarized states in which all 
three spin components were spatially overlapping, since they were 
confined to the same potential trap. However, it is also possible to use 
three different traps, each confining a different component, to initially
separate them, and then allow the system to evolve in the presence of one of
these traps (i.e., turning off the other two). In this section, we present
spin-polarized states, including DW structures, obtained in this way.

\begin{figure}[tbp]
\includegraphics[width=4.1cm,height=3.9cm]{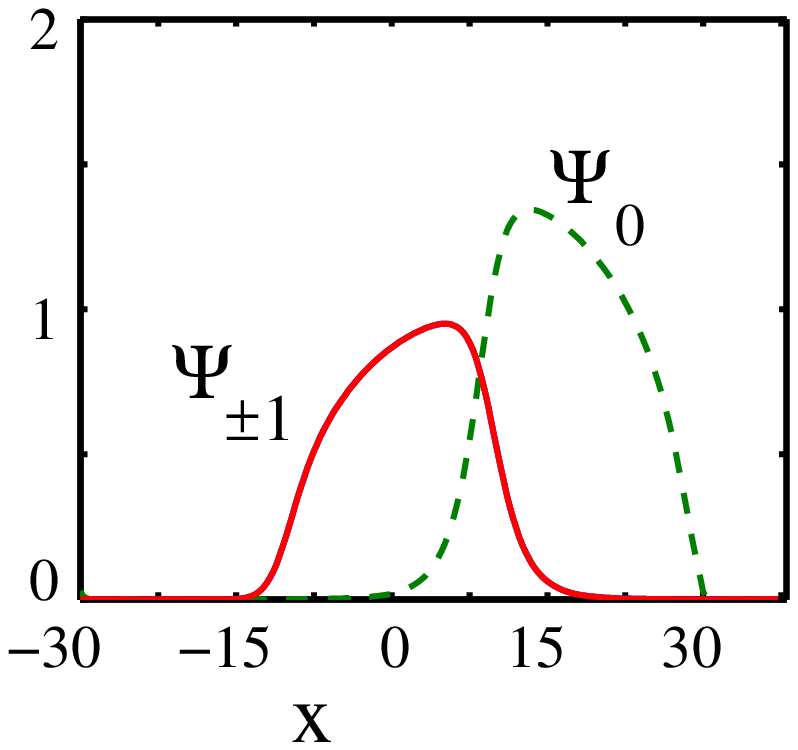} 
~
\includegraphics[width=4.1cm,height=3.67cm]{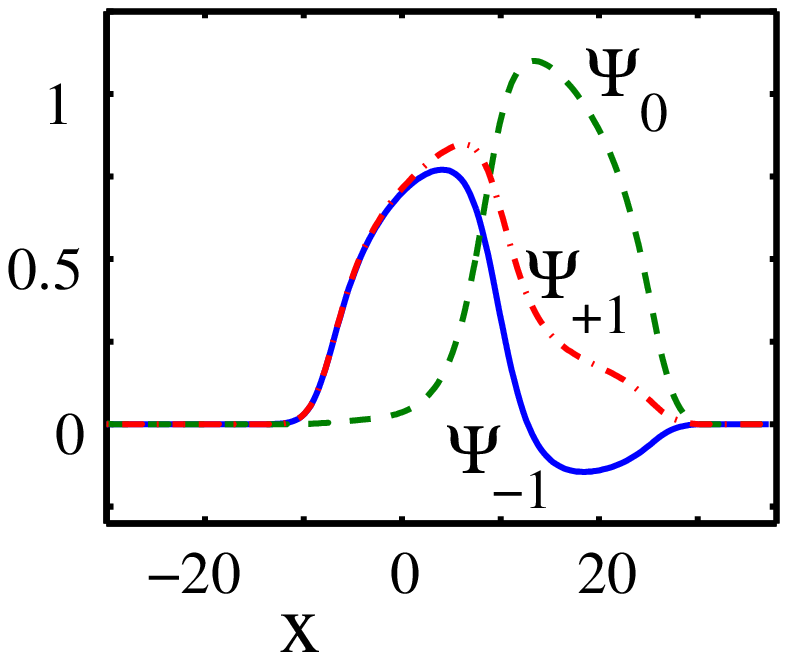}
\includegraphics[width=8.0cm]{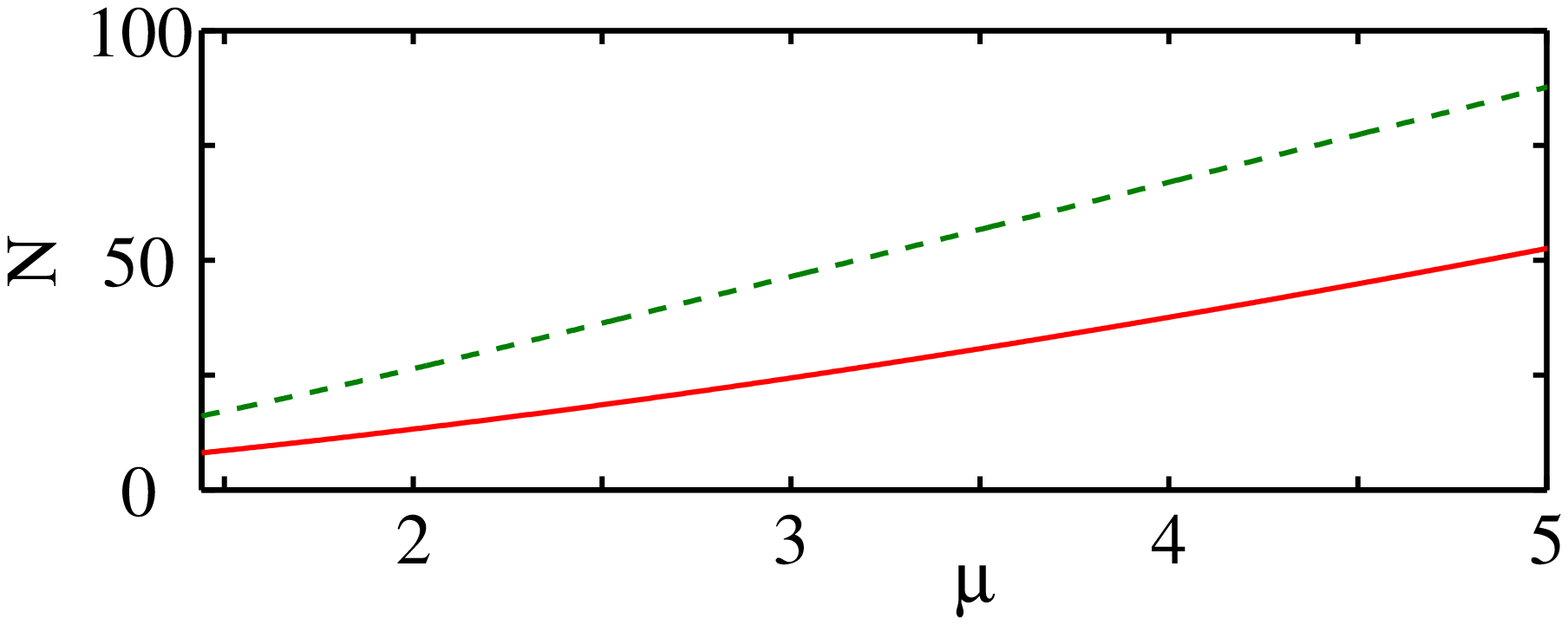}~~ %
\caption{(Color online) 
Top-left panel: the wave functions of the components ($%
\protect\psi _{+1}$ and $\protect\psi _{-1}$ are identical) in a stationary
state found from the initial configuration prepared as shown in Fig.~\protect
\ref{fig7}. 
The resulting spin-polarized state has the form of a domain-wall
like structure between the $\protect\psi _{0}$ and $\protect\psi _{\pm 1}$ 
components.
The parameters are $\Omega =0.1$ and $\mu =2$. 
Top-right panel shows the wavefunctions of the domain-wall state
found at $\protect\mu =1.43$. 
Bottom panel: the norm of each component in the above-mentioned domain-wall
structure versus the chemical potential (the dependences for
$\Delta \protect\theta =0$ or $\protect\pi $ are identical). 
}
\label{fig8}
\end{figure}

First, we describe the initialization of the system. We assume
that the three TF-shaped components 
are initially loaded into three different traps, $V_{j}(x)$, of the same
strength, $\Omega $, centered at different positions:
\begin{equation}
V_{j}(x)=\frac{1}{2}\Omega ^{2}(x-j\Delta x)^{2},~j=-1,0,+1.  \label{tdt}
\end{equation}%
We choose $\Delta x=\Omega ^{-1}$ (i.e., $\Delta x=10$ for $\Omega =0.1$),
which implies initially overlapping TF configurations, see Fig.~\ref{fig7}.
After constructing this state, we turn off the traps $V_{-1}(x)$ and $V_{0}(x)$,
retaining only the rightmost one, $V_{+1}(x)$, which now acts on all the
three components, and feed the initial configuration state into the
fixed-point algorithm, to find new spin-polarized states. Other
possibilities, such as turning off potentials $V_{\pm 1}$ and keeping $V_{0}$%
, arranging the three components in a different way, etc., eventually lead
to retrieving the spin-polarized states presented in the previous
sections, while the approach outlined above [keeping $V_{+1}(x)$ and
switching $V_{-1}(x)$ and $V_{0}(x)$ off] generates new DW patterns, which
are displayed in Fig.~\ref{fig8}, and could not be obtained otherwise. (The
asymmetry of the procedure is instrumental in generating the new states).

The most interesting spin-polarized DW states found following this procedure
correspond to values of the chemical potential $\mu \geq 1.43$ (or norm $%
N\geq 5400$ for $\Omega =0.1$); for smaller values of $\mu$ we typically find structures of the Thomas-Fermi type. 
Two examples, one for $\mu >1.43$,
and another corresponding to 
$\mu =1.43$, are shown in Fig.~\ref{fig8}. In the former one, the $%
\psi _{0}$ component (which has the larger norm)
is centered to the right of the midpoint of the remaining trap ($x_{+1}=10$), while the
identical $\psi _{\pm 1}$ components are pushed to the left, due to the
repulsion from $\psi _{0}$, with a DW created between $\psi _{\pm 1}$ and $\psi _{0}$. Figure \ref{fig8}
displays an example of such a DW structure for 
$\mu =2$ and $\Omega =0.1$.

The state found at the above-mentioned value, $\mu =1.43$, is shown
in the top-right panel of Fig.~\ref{fig8}, for $\Delta \theta =0$. In
this state, the shape of the $\psi _{0}$ component is similar to that
considered above, while the $\psi_{\pm 1}$ components are \emph{not}
identical, in contrast to the above example. In this case the $\psi _{0}$
and $\psi _{\pm 1}$ components overlap over a wider spatial region. Note that $%
\psi _{-1}$ changes its sign at $x\approx 13$, featuring a structure 
resembling the waveform of a dark soliton embedded in a bright one \cite{dark-in-bright}.

\begin{figure}[tbp]
\includegraphics[width=4.0cm,height=3.0cm]{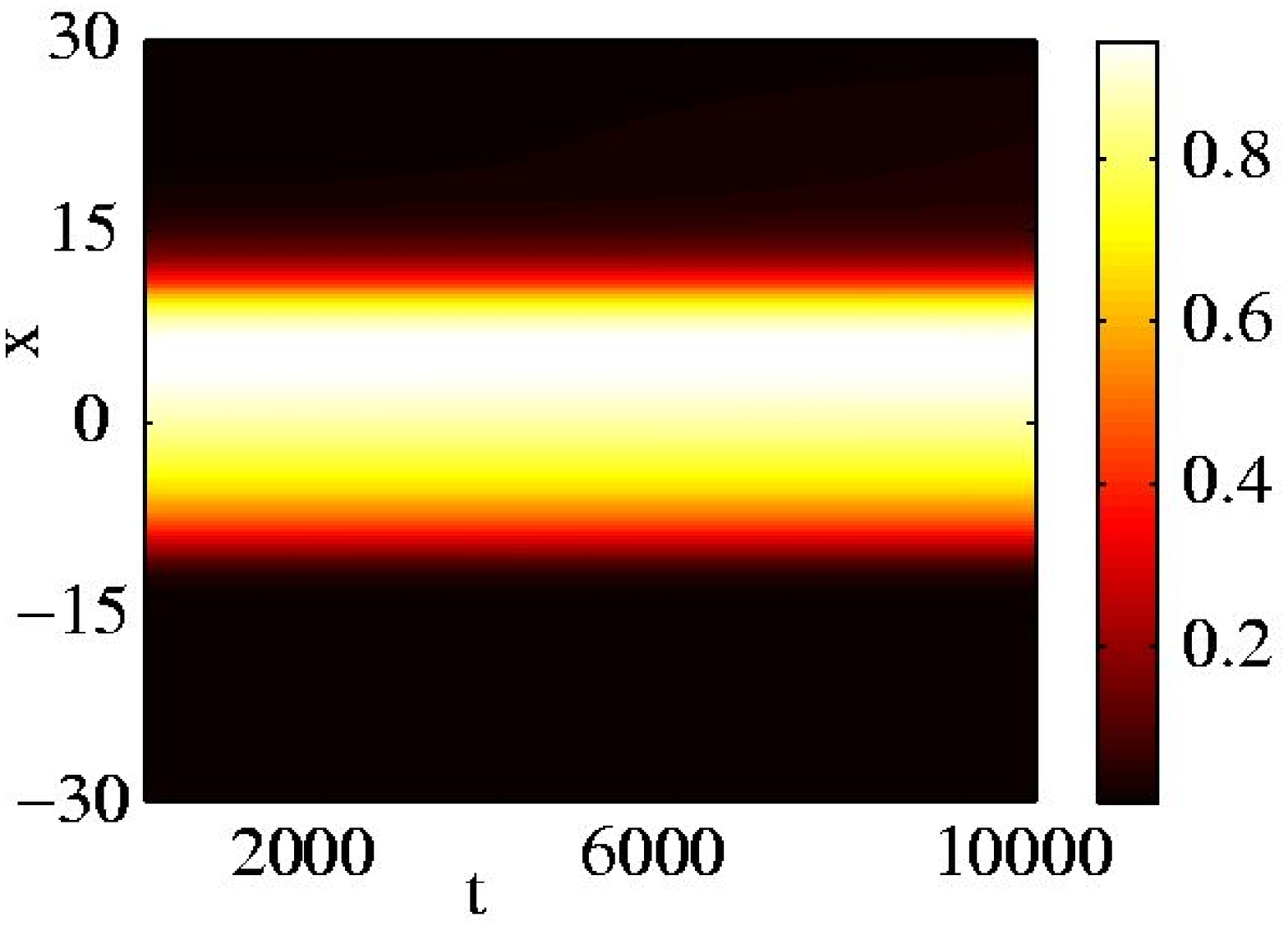} %
~~
\includegraphics[width=4.0cm,height=3.0cm]{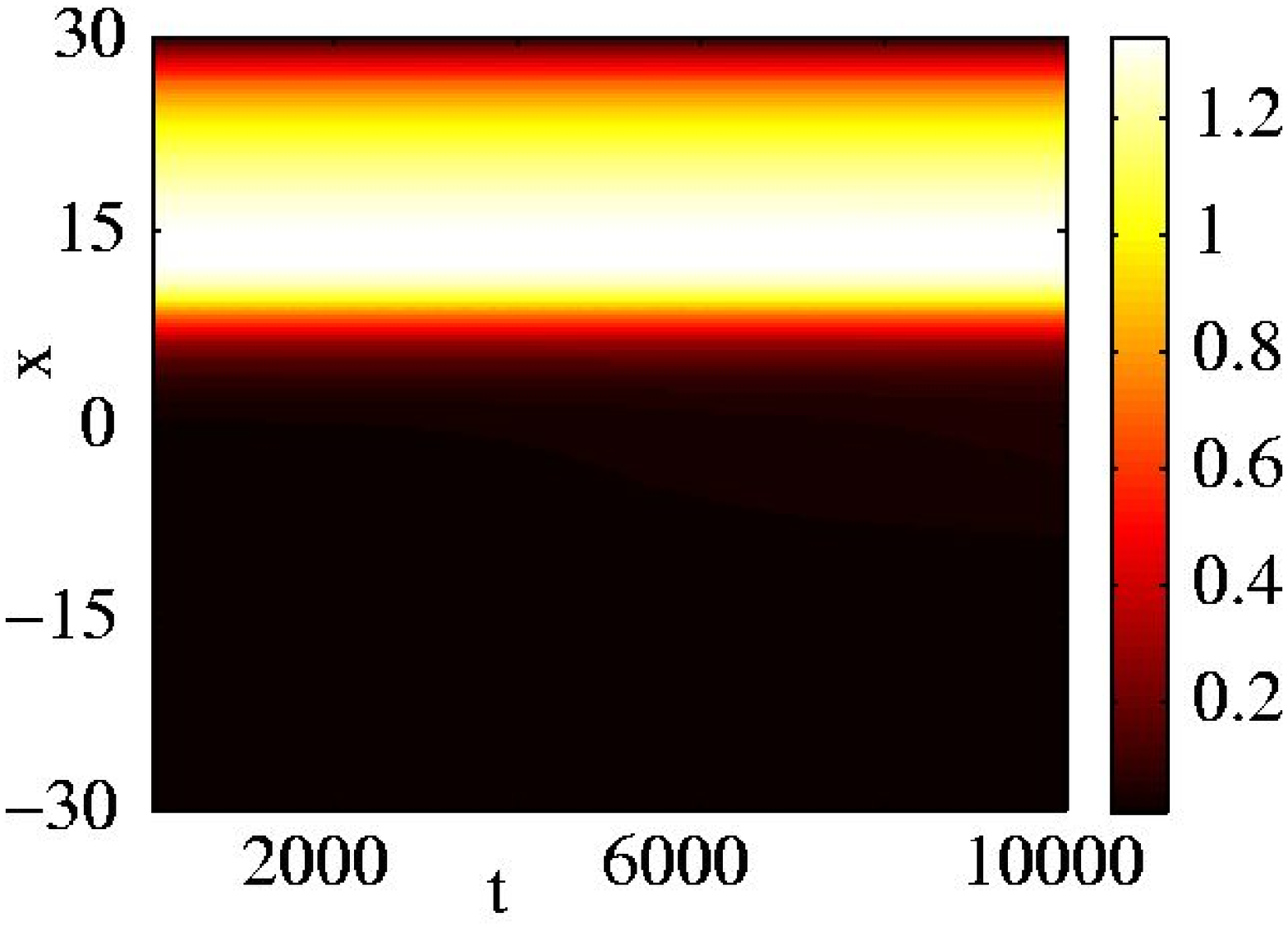}
\caption{(Color online) Evolution of the domain-wall structure
shown in the top-left panel of Fig.~\protect\ref{fig7}, to which a
random perturbation was added. Shown in the left and right panels
are spatiotemporal contour plots of the densities of components
$\protect\psi _{\pm 1}$ (identical ones) and $\protect\psi _{0}$,
respectively.} \label{fig9}
\end{figure}

\begin{figure}[tbp]
\includegraphics[width=2.50cm,height=2.0cm]{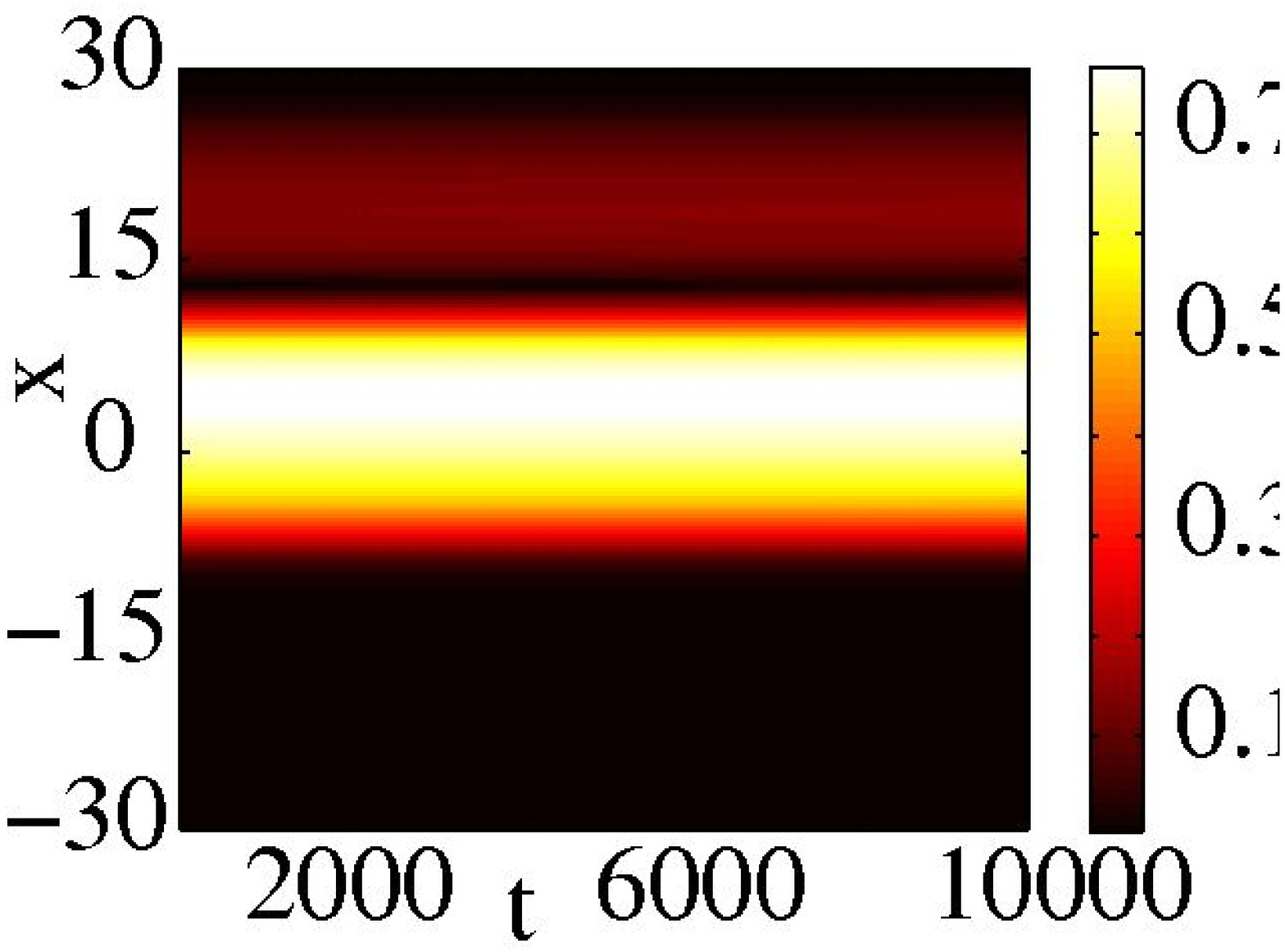} %
~~
\includegraphics[width=2.50cm,height=2.0cm]{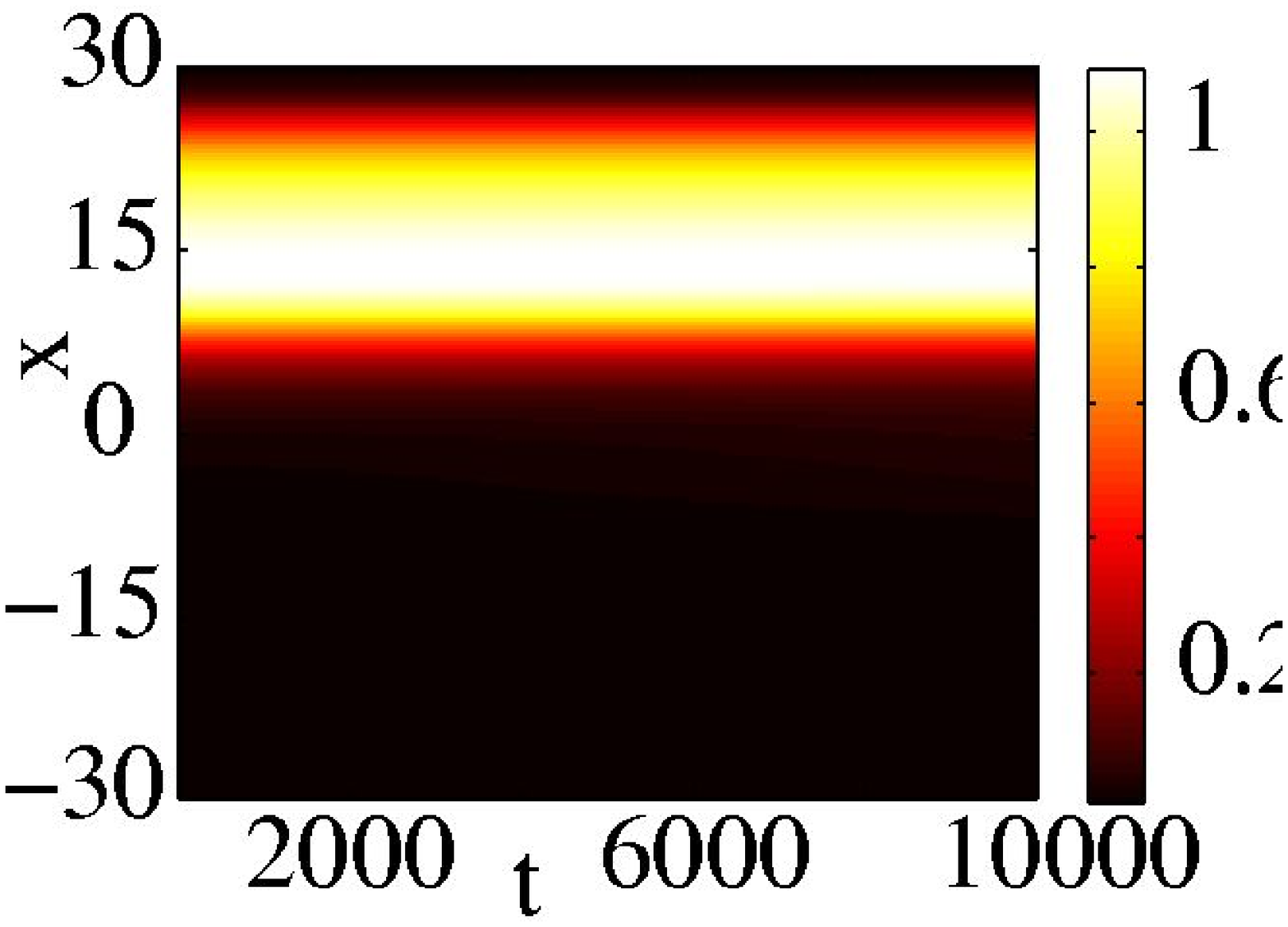} %
~~
\includegraphics[width=2.50cm,height=2.0cm]{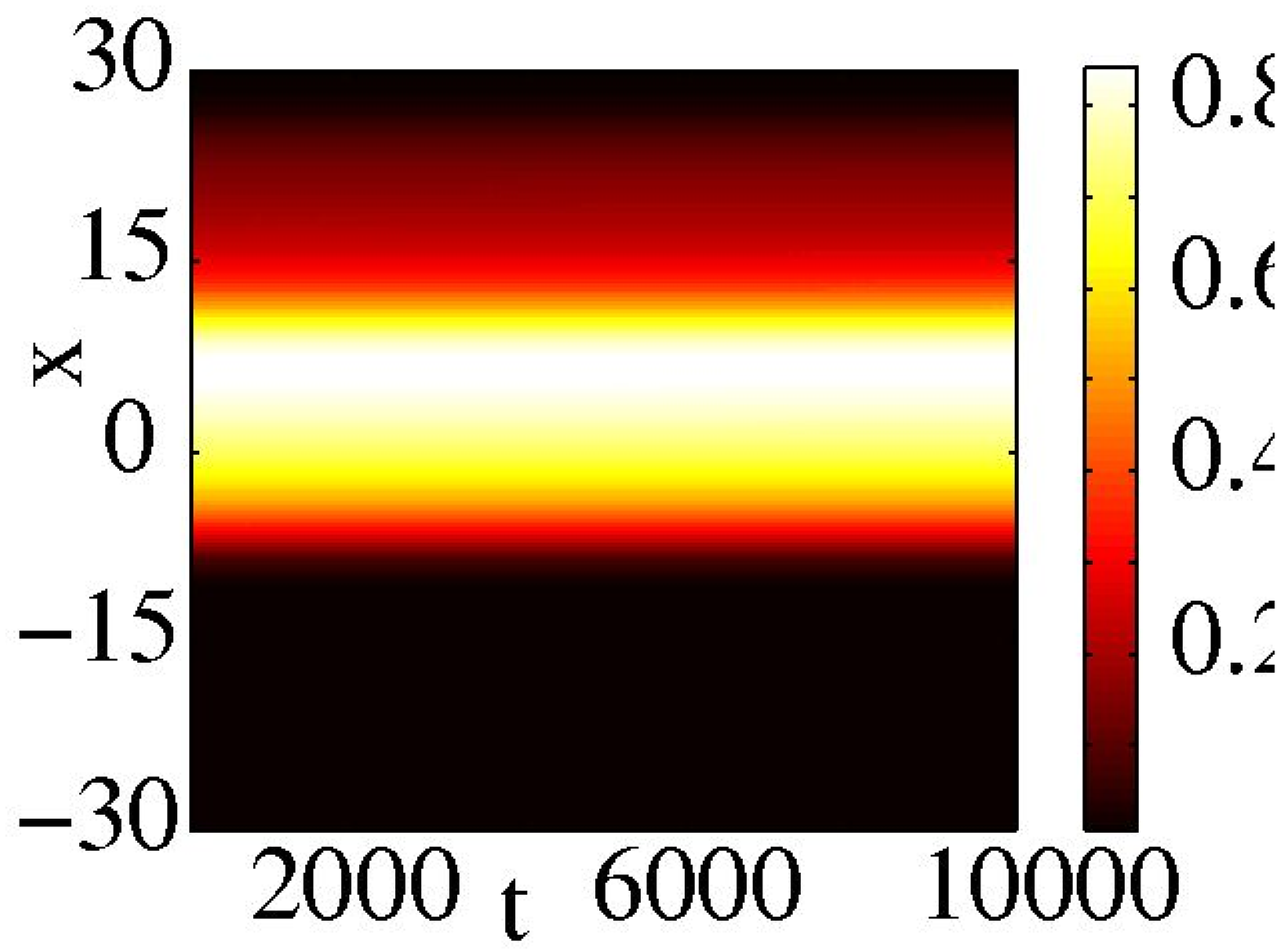}
\caption{(Color online) Same as Fig.~\protect\ref{fig9} but for the state
shown in the top-right panel of Fig.~\protect\ref{fig7}. The left, middle
and right panels show, respectively, the densities of $\protect\psi _{-1}$, $%
\protect\psi _{0}$ and $\protect\psi _{+1}$. Noteworthy is a stationary
dark-soliton-like structure, located at $x\approx 13$ in the $\protect\psi %
_{-1}$ component. }
\label{fig10}
\end{figure}

The stability of the DW states was also investigated by means of the BdG
equations. It was concluded that there are no eigenvalues with a real part
in the interval $1.43\leq \mu \leq 5$, or, equivalently, $5400\leq N\leq 35000$
for $\Omega =0.1$ (not shown here in detail). Thus, the DWs are stable in
this region. Verification of the stability, performed by direct
simulations of Eqs.~(\ref{dvgp1}) and (\ref{dvgp2}), is illustrated in Figs.~%
\ref{fig9} and \ref{fig10}, for $\mu =2$ and $\mu =1.43,$ respectively. It
is obvious that these states are indeed stable, for very large times 
exceeding $t=10000$ (i.e., $12$ seconds in physical units).

\section{Conclusions}

In this work, we have studied spin-polarized states in anti-ferromagnetic
spinor ($F=1$) Bose-Einstein condensates. In particular, our analysis
applies to a quasi-1D spinor condensate of sodium atoms. The considerations 
were based on analytical calculations and numerical computations of the coupled
Gross-Pitaevskii equations for this setting.

Assuming that all three hyperfine (spin) components are confined in the
same harmonic trap, we have found various types of spin-polarized states and
examined their stability. The first family consists of Thomas-Fermi
configurations, considered analytically in the framework of the single-mode
approximation (which assumes the similarity of the spatial profiles of all the
components). Within their existence region, these states were found to be stable. 
Also identified were more complex patterns, 
which include states composed of pulse-like structures in one component,
which induce holes in the other components, and states with holes in all
three components. These states feature windows of weak instability. 
The development of the instability was investigated by means of
direct numerical simulations, which demonstrate that it 
manifests itself at very long times, and results in a weak deformation of the
states with a single hole in some of the components, and a stronger one in the
states with holes in all components.

Fully stable families of spin-polarized states develop from initial
configurations which are initially separated (by means of three different traps) 
components. These states form domain-wall structures between the
components, at values of the chemical potential above a critical value. At
the critical value (corresponding to a certain norm, for a fixed trap's
strength), we have found another spin-polarized state (including a
dark-soliton element), in which all the components partly overlap.

It would clearly be interesting to investigate the existence and stability of
higher-dimensional counterparts of the 1D spin-polarized states found in
this work. Another relevant question for further analysis is whether such
spinor condensates support stable topological objects, such as dark solitons
or vortices. Work in these directions is in progress.

The work of H.E.N. and D.J.F. was partially supported from the Special
Research Account of University of Athens. H.E.N. acknowledges partial
support from EC grants PYTHAGORAS I. P.G.K. acknowledges support from
NSF-CAREER, NSF-DMS-0505663 and NSF-DMS-0619492, as well as the warm
hospitality of MSRI during the initial stages of this work. The work of B.A.M.
was supported, in \ a part, by the Israel Science Foundation through the
Center-of-Excellence grant No. 8006/03, and the German-Israel Foundation
through Grant No. 149/2006. R.C.G. acknowledges support from NSF-DMS-0505663. 
Work at Los Alamos National Laboratory is supported by the USDoE.

\end{document}